\documentclass[floats,floatfix,showpacs,amssymb,prd,twocolumn,superscriptaddress,nofootinbib,nolongbibliography,reprint]{revtex4-1}

\usepackage{amssymb,amsmath,verbatim,mathtools,needspace,enumitem,etoolbox,graphicx,physics,microtype,afterpage,xspace,tabularx,lmodern,multirow,mathrsfs}
\usepackage{gensymb}
\usepackage[normalem]{ulem}
\usepackage[dvipsnames, usenames]{xcolor}
\definecolor{linkcolor}{rgb}{0.0,0.3,0.5}
\usepackage[unicode, colorlinks=true, linkcolor=linkcolor, citecolor=linkcolor, filecolor=linkcolor, urlcolor=linkcolor, linktocpage, breaklinks]{hyperref}
\usepackage[all]{hypcap}
\usepackage{subfigure}
\usepackage[T1]{fontenc}
\usepackage[utf8]{inputenc}
\usepackage[usenames,dvipsnames]{xcolor}
\usepackage{enumerate}

\hypersetup{colorlinks=true,citecolor=romared,linkcolor=romared,urlcolor=romared}

\definecolor{romared}{RGB}{142,0,28}
\newcommand{\nn}{\nonumber}

\newcommand{\be}{\begin{equation}}
\newcommand{\ee}{\end{equation}}
\newcommand{\beq}{\begin{eqnarray}}
\newcommand{\eeq}{\end{eqnarray}}

\usepackage{aas_macros}
\usepackage{makecell}
\usepackage{soul}
\usepackage{amssymb}
\usepackage{longtable}

\usepackage{lipsum}

\newcolumntype{Y}{>{\centering\arraybackslash}X}

\def\scri{\mathscr{I}}

\begin{document}

\title{Limiting geometry and spectral instability in Schwarzschild--de Sitter spacetimes}

\begin{abstract}
We revisit the quasinormal mode (QNM) problem in Schwarzschild--de Sitter spacetimes providing a unified infrastructure tailored for studying limiting configurations. Geometrically, we employ the hyperboloidal framework to explicitly implement Geroch's rigorous limiting procedures for families of spacetimes. This enables a controlled transition between Schwarzschild, de Sitter, and Nariai geometries. Numerically, we introduce the analytical mesh refinement technique into quasinormal mode calculations, successfully recovering---within the appropriate limiting scenarios---both known families of quasinormal modes: complex light ring modes and purely imaginary de Sitter modes.
We interpret these results in terms of spectral instability, where the notions of stable and unstable modes depends on the specific spacetime limit under consideration. In the Schwarzschild limit, de Sitter modes appear as a destabilizing effect on the continuous branch cut at $\omega = 0$. Conversely, the branch cut can be understood as emerging from an infinite accumulation of discrete modes at $\omega = 0$ in the transitional regime. We propose a heuristic measure of QNM density to characterize this accumulation and highlight the need for a more rigorous study of potential branch cut instabilities---especially relevant in the context of late-time gravitational wave signals.
The proposed infrastructure provides a general and extensible framework for investigations in more complex spacetimes, such as Reissner--Nordström--de Sitter or Kerr--Newman--de Sitter.
\end{abstract}

\author{Yi Zhou}
\affiliation{Center of Gravity, Niels Bohr Institute, Blegdamsvej 17, 2100 Copenhagen, Denmark}

\author{Rodrigo Panosso Macedo}
\affiliation{Center of Gravity, Niels Bohr Institute, Blegdamsvej 17, 2100 Copenhagen, Denmark}

\maketitle

\section{Introduction}
Black hole (BH) quasinormal modes (QNMs) are characteristic frequencies associated with BH spacetimes. They naturally emerge when Einstein’s equations are linearized around a fixed background geometry and the resulting linear equations are solved to model a physical scenario in which energy flows into the black hole and is radiated outward into the wave zone~\cite{Chandrasekhar:1985kt,Kokkotas:1999bd,Nollert:1999ji,Berti:2009kk,Konoplya:2011qq,Berti:2025hly}. Encoding detailed information about the geometry, QNMs are a cornerstone of gravitational wave astronomy through the so-called BH spectroscopy program~\cite{Dreyer:2003bv,Berti:2005ys,LISAConsortiumWaveformWorkingGroup:2023arg,Berti:2025hly}. In fact, the dominant QNM has been measured in several gravitational wave (GW) events~\cite{LIGOScientific:2016aoc,LIGOScientific:2020tif,LIGOScientific:2021sio}, with increasing efforts devoted to identifying subdominant modes~\cite{Isi:2019aib,Capano:2020dix,Capano:2021etf,Cotesta:2022pci,Capano:2022zqm,Forteza:2022tgq,Finch:2022ynt,Abedi:2023kot,Carullo:2023gtf,Baibhav:2023clw,Nee:2023osy,Zhu:2023mzv,Siegel:2023lxl,Gennari:2023gmx}.

Beyond the observational challenges in BH spectroscopy, the field also faces theoretical advances. In particular, the phenomenon of QNM spectral instability has gained increasing attention in recent years~\cite{Aguirregabiria:1996zy,Vishveshwara:1996jgz,Nollert:1996rf,Daghigh:2020jyk,Qian:2020cnz,Jaramillo:2020tuu,Jaramillo:2021tmt,Cheung:2021bol,Cardoso:2024mrw}. In short, there is growing heuristic and mathematical evidence that small perturbations in the wave operators governing the linear dynamics can lead to large deviations in the QNM spectrum. Open questions remain regarding the physical significance of these instabilities~\cite{Jaramillo:2021tmt,Cardoso:2024mrw}, their possible imprint in time-domain signals~\cite{Jaramillo:2021tmt,Spieksma:2024voy}, and the mathematical foundations of the phenomenon~\cite{Jaramillo:2020tuu,Boyanov:2023qqf,Besson:2024adi,Carballo:2025ajx}.

Underpinning these theoretical advances, the hyperboloidal framework has established itself as a foundational strategy in BH perturbation theory~\cite{Zenginoglu:2011jz,PanossoMacedo:2023qzp,PanossoMacedo:2024nkw}. This framework provides not only a geometric approach for treating the asymptotic regions of spacetime, but also a robust numerical infrastructure for solving the resulting singular equations~\cite{PanossoMacedo:2014dnr,Jaramillo:2020tuu,PanossoMacedo:2022fdi,Bourg:2025lpd,Assaad:2025nbv}. In light of these recent developments in the field, this work revisits the problem of wave propagation in the Schwarzschild--de Sitter (SdS) spacetime, with particular emphasis on the distinct limiting geometries arising from the SdS solution.

The SdS spacetime describes a nonspinning, noncharged black hole embedded in an asymptotically de Sitter universe; i.e., it is a black hole solution of Einstein’s equations with a positive cosmological constant, $\Lambda > 0$. Its astrophysical relevance stems from cosmological observations indicating the existence of a positive cosmological constant. As a result, astrophysical black holes are immersed in spacetimes with a small but nonzero cosmological constant, which modifies the asymptotic structure of the modeling spacetime on large scales. 

In the era of GW astronomy---particularly with an eye toward next-generation detectors---there has been growing theoretical interest in the late-time behavior of GW signals from binary black hole mergers~\cite{Blanchet:1994ez,DeAmicis:2024eoy,Cardoso:2024jme,Ma:2024hzq}. Since this late-time behavior is strongly influenced by the spacetime’s asymptotic structure, it is important to understand the time evolution of perturbations in this regime for fields propagating on an SdS background~\cite{Brady:1996za,Konoplya:2024ptj,Correa:2024xki}. Of particular interest in the astrophysical context is the regime $\Lambda \gtrsim 0$. 

Physically, this regime can be interpreted as a small modification by a positive cosmological constant $\Lambda$ into a Schwarzschild spacetime with BH mass $M$. An alternative small parameter representation is to consider a small BH with mass $M\gtrsim 0$ embedded in an otherwise de Sitter universe. Given the scale invariance of general relativity, these two scenarios correspond to different limits of one single dimensionless constant $\sim M \sqrt{\Lambda}$. Thus, the SdS solution is constituted, in fact, by a one-parameter family of spacetimes.

Formally, the mathematical framework to rigorously study limiting procedures for families of spacetimes was developed by Geroch~\cite{Geroch:1969ca}. In broad terms, the idea is to embed the entire family of four-dimensional spacetimes into a higher-dimensional manifold, where the additional dimension tracks the parameter labeling the family. However, this extension is not unique: to define a limit, one must specify a prescription for identifying spacetime points across different members of the family. Different identification schemes may yield different limiting spacetimes, even for the same one-parameter family. 

At the level of the QNMs that characterize the spacetime geometry, one also encounters key distinctions between the Schwarzschild and de Sitter spacetimes. The de Sitter spacetime possesses a discrete set of purely imaginary modes~\cite{Lopez-Ortega:2006aal}, while the Schwarzschild spacetime features a discrete set of complex QNMs associated with the presence of a light ring, as well as a continuous spectrum represented by a branch cut along the negative imaginary axis~\cite{Leaver:1986gd}. Consequently, the QNM spectrum of the Schwarzschild--de Sitter spacetime exhibits a hybrid structure, comprising both complex (light ring) modes and purely imaginary (de Sitter--like) modes~\cite{Konoplya:2004uk,Cardoso:2017soq,Jansen:2017oag,Konoplya:2022xid}.

From a technical perspective, traditional methods based on Leaver’s algorithm~\cite{leaver1985analytic} provide useful tools for computing QNMs in SdS spacetime~\cite{Cardoso:2017soq,Konoplya:2022xid,Stuchlik:2025mjj}. A more systematic study of the two families of QNMs in SdS spacetime has also been carried out using intricate numerical techniques based on the so-called Bernstein polynomials~\cite{Fortuna:2020obg,Konoplya:2022xid}. These polynomials form a set of nonorthogonal basis functions, introduced to address the challenges posed by the outgoing boundary conditions characteristic of the QNM problem. 

It is important to note, however, that in both approaches the boundary conditions are first treated analytically, through the introduction of auxiliary regular functions defined at the level of the frequency-domain formulation involving radial ordinary differential equations. Solutions are then obtained by applying either a Taylor expansion ansatz or spectral methods, in order to handle the resulting singular equations for the auxiliary functions. In fact, it is now understood that these auxiliary functions correspond to the frequency-domain representation of the hyperboloidal framework~\cite{Zenginoglu:2011jz,PanossoMacedo:2023qzp,PanossoMacedo:2024nkw}, particularly within the so-called minimal gauge~\cite{Ansorg:2016ztf,PanossoMacedo:2018hab,PanossoMacedo:2019npm,PanossoMacedo:2023qzp,PanossoMacedo:2024nkw}.

By revisiting the QNM problem on the SdS spacetime from a hyperboloidal formulation, we unify the various aspects discussed above under a single, coherent approach. As expected, the introduction of hyperboloidal coordinates---designed to geometrically access the relevant asymptotic regions---resolves the typical shortcomings of the traditional formulation, where QNM eigenfunctions are ill defined and diverge at the horizons. The framework offers a direct spacetime-based interpretation~\cite{Zenginoglu:2011jz,PanossoMacedo:2023qzp,PanossoMacedo:2024nkw} for analytical manipulations that are otherwise carried out solely at the level of the radial frequency-domain equations~\cite{Jansen:2017oag,Konoplya:2022xid}. Most importantly for this work, we show that the hyperboloidal formulation provides a natural geometric implementation of Geroch’s limiting procedure.

Once the geometric foundations are established, the technical difficulties that hinder QNM analyses in the limiting regime become more transparent. We address these challenges using spectral methods based on the well-known Chebyshev polynomials, in contrast to more elaborate techniques involving Bernstein polynomials. To manage the numerical difficulties that arise in delicate regions of the parameter space, we introduce the so-called analytical mesh refinement (AnMR) technique~\cite{Meinel:2008kpy,Pynn:2016mtw,PanossoMacedo:2022fdi} into the context of QNM computations. 

Finally, our results are interpreted in the context of QNM spectral instability~\cite{Jaramillo:2020tuu,Cardoso:2024mrw}. In contrast to previous studies~\cite{Sarkar:2023rhp,Destounis:2023nmb} that examined the spectral instability of the SdS QNM spectrum directly, the objective of this work is to follow the perspective of Ref.~\cite{Cardoso:2024mrw}, interpreting the emergence of new families of QNMs as a manifestation of spectral instability triggered by the introduction of additional physical length scales into the system.

This work is organized as follows. Section~\ref{sec:SdS} reviews the SdS spacetime and the propagation of a massless scalar field on this background, treated as a probe field. Section~\ref{sec:Hyp} introduces the hyperboloidal framework for the spacetime and formulates the QNM eigenvalue problem. Section~\ref{sec:limits} is devoted to a detailed discussion of the spacetime limits. The numerical methods are presented in Sec.~\ref{sec:NumMeth}, with particular emphasis on analytical mesh refinement. Section~\ref{sec:Results} presents the results, including convergence tests used to benchmark the new numerical techniques, and an analysis of the QNM spectrum in the context of spectral instability. We conclude in Sec.~\ref{sec:Conclusion}. Throughout this work, we use geometrized units with $G = c = 1$.

\section{Schwarzschild de Sitter spacetime}\label{sec:SdS}

We start with the line element of SdS spacetime. In Schwarzschild coordinates $(t, r, \theta, \varphi)$ the line element reads
\begin{equation}
	\label{eq:line_element}
    d s^2 = -f(r) dt^2 + f^{-1}(r) dr^2 + r^2 d\varpi^2 ,
\end{equation}
with $ d\varpi^2 = d\theta^2 + \sin^2\theta d\varphi^2 $ the metric of the two-sphere.
The function $f(r)$ is parametrized either by the spacetime mass $M$ and cosmological constant $\Lambda$, or by its real and positive roots $r_\Lambda \ge r_H \ge 0 $, with  $r_H$ representing the black hole horizon and $r_\Lambda$ the cosmological horizon. Specifically, the metric function reads
\begin{equation}
    \label{eq:f_of_r}
    \begin{aligned}
        f(r) & = 1 - \frac{2M}{r} - \frac{\Lambda}{3} r^2 \\
             & = - \frac{\Lambda}{3} r^2
                 \left( 1- \frac{r_H}{r} \right)
                 \left( 1- \frac{r_\Lambda}{r} \right)
                 \left( 1- \frac{r_o}{r} \right),
    \end{aligned}
\end{equation}
with $r_o = - (r_H + r_\Lambda)$ the third root of $f(r)$. It is also convenient to define the tortoise coordinate $r_*(r)$ via $ {d r_*}/{dr} = {1}/{f(r)}$, which integrates to 
\beq
    \label{eq:tort_def}
    r_* &=& \dfrac{3 r_H}{\Lambda (r_\Lambda - r_H)(r_H-r_o)} \ln\left| 1- \dfrac{r_H}{r}\right| \nn \\
    &+& \dfrac{3 r_\Lambda}{\Lambda ( r_H- r_\Lambda)(r_\Lambda-r_o)} \ln\left|  1- \dfrac{r_\Lambda}{r}\right| \nn \\
    &+& \dfrac{3 r_\Lambda}{\Lambda ( r_H - r_o)(r_o- r_\Lambda)} \ln\left|  1- \dfrac{r_o}{r}\right|.
\eeq

From Eq.~\eqref{eq:f_of_r}, one directly derives the relation between the parameters $\Lambda $ and $ M $ and the horizons via
\begin{gather}
    \Lambda = \frac{3}{ r_H^2 + r_H r_\Lambda + r_\Lambda^2 } , \\
    M = \frac{ r_H r_\Lambda \left( r_H + r_\Lambda \right) }{2\left( r_H^2 + r_H r_\Lambda + r_\Lambda^2 \right)} .
\end{gather}
Alternatively, Cardano's formula for a cubic equation gives for $ \phi = \arccos ( 3 M \sqrt{\Lambda} )$
\beq
\label{eq:rh_def}
r_H  = \dfrac{1}{\sqrt{\Lambda}} \left(  \cos \left(\dfrac{\phi}{3}\right) - \sqrt{3} \sin \left(\dfrac{\phi}{3}\right)     \right), \\
\label{eq:rlambda_def}
r_\Lambda  = \dfrac{1}{\sqrt{\Lambda}} \left(  \cos \left(\dfrac{\phi}{3}\right) + \sqrt{3} \sin \left(\dfrac{\phi}{3}\right)     \right).
\eeq
It is evident that the horizons' coordinate location changes with respect to the parameters $M$ and $\Lambda$, i.e., $r_H(M, \Lambda)$ and $r_\Lambda(M, \Lambda)$. As expected, Eqs.~\eqref{eq:rh_def} and \eqref{eq:rlambda_def} yield
\beq
\label{eq:SchwLimit}
&&\lim_{\Lambda \rightarrow 0^+ } r_H(M, \Lambda) = 2 M, \,\, \lim_{\Lambda \rightarrow 0^+} r_\Lambda(M, \Lambda) = +\infty, \\
\label{eq:dSLimit}
&&\lim_{M \rightarrow 0 } r_H(M, \Lambda) = 0, \,\, \lim_{M \rightarrow 0} r_\Lambda(M, \Lambda) = \sqrt{\dfrac{3}{\Lambda}}.
\eeq 
From the two characteristic length scales provided by the horizons, a description of SdS geometry as an one-parameter family of solution follows via the variable
\begin{equation}
    \label{eq:eta_def}
    \eta = \dfrac{r_H}{r_\Lambda}.
\end{equation}
Even though $\eta\in[0,1]$, there are actually {\em four} different limits of the spacetime to be considered for a geometrical completeness. 

The limit $\eta \rightarrow 0$ corresponds either to the Schwarzschild limit via Eqs.~\eqref{eq:SchwLimit} or de Sitter limit via Eqs.~\eqref{eq:dSLimit}. Moreover, the limit $\eta\rightarrow 1$ corresponds either to the direct extremal SdS configuration~\cite{Podolsky:1999ts}, or the Nariai solution~\cite{1950SRToh..34..160N}, akin to the near-horizon geometry of the extremal spacetimes~\cite{Carroll:2009maa,Kunduri:2013gce}. The direct extremal limit, however, has an important difference with respect to all other configurations and limits. In this case, the Killing vector $\partial_t$ is always spacelike and the spacetime does not contain any ``exterior'' region where static geodesic observers exist. Thus, a Fourier study of the wave equation on the direct extremal SdS spacetime is not possible. 

Indeed, for configurations where the spacetime admits a region with a timelike Killing vector, one may consider a massless scalar field $\Psi$ as a proxy observable capturing the key qualitative properties of perturbations propagating on the SdS spacetime. The scalar field's dynamics are dictated by Klein-Gordon equation $\square \Psi = 0$, and  a standard separation of variables in terms of the spherical harmonics  $Y_{\ell m}(\theta, \varphi)$  and a Fourier decomposition via 
$
\Psi(t,r,\theta, \varphi) = r^{-1} \sum_{\ell,m}  e^{-i\omega t}\psi_\ell(r) Y_{\ell m}(\theta, \varphi),
$
yields 
\beq
\label{eq:WaveEq}
\psi_{\ell}{}_{,r_*r_*} - \left( V_\ell - \omega^2 \right) \psi_{\ell} = 0,
\eeq
with the potential
\beq
V_\ell(r) = f(r) \left( \dfrac{\ell(\ell+1)}{r^2} + \dfrac{f'(r)}{r} \right).
\eeq
The boundary conditions 
\beq
\label{eq:BC}
\psi_{\ell} \sim e^{\pm i\omega r_*}, \quad r_* \rightarrow \pm \infty
\eeq
ensure that energy flows into the black hole at $r_H$ and out across the cosmological horizon $r_\Lambda$, with the QNM frequencies the values $\omega_{n}$ for which the solution satisfies Eq.~\eqref{eq:BC}.

In the next section we discuss the SdS spacetime from the perspective of the hyperboloidal framework. Not only does this strategy provide a natural geometrical approach to incorporate the boundary conditions~\eqref{eq:BC}, but the hyperboloidal framework also offers a clean route to describe the three spacetime limits within the geometrical prescription by Geroch \cite{Geroch:1969ca}, as discussed in Sec.~\ref{sec:hyp_coord} and \ref{sec:limits}.

\section{Hyperboloidal Coordinates}\label{sec:hyp_coord}\label{sec:Hyp}

A generic transformation into hyperboloidal coordinates $(\tau, \sigma, \theta, \varphi)$ reads \cite{Zenginoglu:2011jz}  in the notation introduced in Ref.~\cite{PanossoMacedo:2023qzp},
\begin{equation}
    \label{eq:hyp_coord}
    t = \lambda \Big( \tau - H(\sigma) \Big), \quad r = \lambda \dfrac{\rho(\sigma)}{\sigma},
\end{equation}
The constant $\lambda$ is a typical length scale of the spacetime. As we will discuss, it plays an important role according to the particular limit under consideration. 

The height function $H(\sigma)$ and radial function $\rho(\sigma)$ represent degrees of freedom fixing the hyperboloidal slice $\tau =$ constant, and the radial compactification. The radial compactification can also be represented in terms of the dimensionless tortoise coordinate
\begin{equation}
    x(\sigma) = \dfrac{r_* (r(\sigma))}{\lambda}.
\end{equation}
From Eq.~\eqref{eq:tort_def}, we observe that the dimensionless tortoise coordinate assumes the form~\cite{PanossoMacedo:2023qzp}
\begin{equation}
    \label{eq:tort_x}
    x(\sigma) = x_H(\sigma) + x_\Lambda(\sigma) + x_o(\sigma),
\end{equation}
with the functions $x_H(\sigma)$ and $x_\Lambda(\sigma)$  singular at the black hole $\sigma_H$ and cosmological $\sigma_\Lambda$ horizons, respectively. The function $x_o(\sigma)$, on the other hand, is regular in the entire radial domain.

Under the transformation \eqref{eq:hyp_coord}, the line element \eqref{eq:line_element} conformally rescales as
\beq
d\bar s^2 &=& \sigma^2 d s^2 \nn\\
\label{eq:conf_line_element}
&=&\lambda^2\beta(\sigma) \bigg( -p(\sigma) d\tau^2  + 2 \gamma(\sigma) + w(\sigma) d\sigma^2  \bigg) \nn \\
&&+ \lambda^2\rho(\sigma)^2 d\varpi^2,
\eeq
with the hyperboloidal metric functions given by~\cite{PanossoMacedo:2023qzp}
\beq
\label{eq:hyp_line_func1}
& \beta(\sigma) = \rho(\sigma) + \sigma \rho'(\sigma), \quad p(\sigma) = -\dfrac{1}{x'(\sigma)} \\
\label{eq:hyp_line_func2}
& \gamma = p(\sigma) H'(\sigma), \quad w(\sigma) = \dfrac{1-\gamma(\sigma)^2}{p(\sigma)}.
\eeq
The conformal rescaling in Eq.~\eqref{eq:conf_line_element} differs slightly from the one suggested in Ref.~\cite{PanossoMacedo:2023qzp}. Here, we explicitly retain the length scale $\lambda$ in Eq.~\eqref{eq:conf_line_element}, keeping the conformal line element a quantity 
with dimension $[ds^2] = ({\rm Length})^2$. This choice will play an important role when studying the extremal limit $\eta\rightarrow 1$ in Sec.~\ref{sec:Nariai_limit}.

To fix the hyperboloidal degrees of freedom, we restrict ourselves to the minimal gauge class\cite{PanossoMacedo:2023qzp}. 

\subsection{The minimal gauge}\label{sec:minimal gauge}
To simplify the differential $dr = - \beta(\sigma)\sigma^{-2} d\sigma$, the minimal gauge~\cite{PanossoMacedo:2023qzp} fixes the radial transformation by imposing $\beta(\sigma)=$ constant, which implies
\begin{equation}
    \label{eq:def_rho}
    \rho(\sigma) = \rho_0 + \rho_1 \sigma.
\end{equation}
The free parameters $\rho_0$ and $\rho_1$ allow us to map specific spacetime hypersurfaces into surfaces at a fixed $\sigma$ value. This property provides us with a practical way of implementing Geroch~\cite{Geroch:1969ca} geometrical limiting procedure to study the two possible limits $\eta\rightarrow 0$, and the extremal limit $\eta\rightarrow 1$. 

More specifically, there are four spacetime surfaces of particular importance: the singularity $r=0$, horizons $r=r_H$ and $r_\Lambda$, and the asymptotic region $r\rightarrow \infty$. Equation~\eqref{eq:hyp_coord} maps them from the set $\{ 0, r_H, r_\Lambda, \infty \}$ into $\{ \sigma_\text{sing}, \sigma_H, \sigma_\Lambda, 0 \}$. 

Recall that in the original Schwarzschild coordinates, the horizons $r_H(M,\Lambda)$ and $r_\Lambda(M,\Lambda)$ coordinate locations depend parametrically on the mass and cosmological constant, whereas  the singularity and the asymptotic region are fixed, respectively, at $r=0$ and $r\rightarrow \infty$, regardless of $M$ and $\Lambda$. 

In the minimal gauge hyperboloidal coordinates, we can choose which surface remains fixed at a coordinate location, and which has a coordinate value depending parametrically on $\eta$. 

The asymptotic region is, by construction, fixed at a coordinate ($\sigma = 0$), completely independent from the choice of $M$ and $\Lambda$. However, the parameters $\rho_0$ and $\rho_1$ offer a freedom to place two of the remaining relevant surfaces at fixed values. As detailed in Sec.~\ref{sec:limits}, this freedom is essential to ensure the correct spacetime limits as $\eta \rightarrow 0$ or  $\eta \rightarrow 1$.

To fix the height function in Eq.~\eqref{eq:hyp_coord}, we recall that, as $\sigma \rightarrow \sigma_H$, the hyperboloidal hypersurface $\tau=$constant must behave as the ingoing null coordinate $\tau \sim v = t + r_* $, whereas for $\sigma \rightarrow \sigma_\Lambda$, the surface must behave as the outgoing null coordinate $\tau \sim u = t - r_* $. A straightforward way to ensure these properties is by reversing the sign of $x_\Lambda(\sigma)$ in \eqref{eq:tort_x}, i.e., $H(x) = x_H(\sigma) - x_\Lambda(\sigma) + x_o(\sigma)$. This line of reasoning is in accordance with the in-out strategy~\cite{PanossoMacedo:2023qzp} and it fixes a hyperboloidal coordinate system for $\eta \neq 0$. 

As we will show, however, this strategy {\em does not yield a well-defined limit} to the Schwarzschild geometry as $\eta \rightarrow 0$. Instead, one needs to resort to the out-in strategy~\cite{PanossoMacedo:2023qzp}. In practical terms, this approach amounts to also reversing the sign of the regular term $x_o(\sigma)$ in \eqref{eq:tort_x}, i.e., the height function is given by
\begin{equation}
	\label{eq:H_def}
    H(\sigma) = x_H(\sigma) - x_\Lambda(\sigma) - x_o(\sigma).
\end{equation}

With expressions \eqref{eq:def_rho} and \eqref{eq:H_def} fixing the hyperboloidal transformation \eqref{eq:hyp_coord} in the minimal gauge, one can study the limits of the SdS geometry with respect to the parameter $\eta$. 

The only remaining parameter in Eq.~\eqref{eq:hyp_coord} is the typical length scale $\lambda$, which can be associated either with the horizon $r_H$ or the cosmological length $r_\Lambda$.  Thus, the choice of $\lambda$  plays an important role in the limiting process, as well. In particular, fixing the unit of length also affects how dimension observables, such as the QNM frequencies $\omega_n$ scale.

\subsection{Quasinormal modes}
Since the time slices $\tau=$ constant penetrate the black hole and cosmological horizons, the boundary conditions \eqref{eq:BC} are automatically satisfied~\cite{Zenginoglu:2011jz,PanossoMacedo:2023qzp,PanossoMacedo:2024nkw}. Indeed, as a direct consequence of the coordinate transformation \eqref{eq:hyp_coord} the scalar field $\psi_\ell(r)$ transforms as\footnote{The relation $r(\sigma)$ is assumed in Eqs.~\eqref{eq:psi_hyp} and \eqref{eq:hyp_pot}. }
\beq
\label{eq:psi_hyp}
\psi_\ell(r) = Z(\sigma) \bar{\psi_\ell}(\sigma), \quad Z(\sigma) = e^{s H(\sigma)},
\eeq
with $Z(\sigma)$ responsible for ensuring the behavior in Eq.~\eqref{eq:BC}~\cite{Zenginoglu:2011jz,PanossoMacedo:2023qzp}. The dimensionless frequency $s$ is fixed by the length scale $\lambda$ via
\beq
\label{eq:QNM_lambda}
\lambda \omega = i s.
\eeq
As already mentioned, fixing $\lambda$ with respect to the black hole or cosmological horizon scales will impact how the QNMs behave in the limiting process $\eta\rightarrow 0$ or $\eta \rightarrow 1$.

In terms of the fields $\bar{\psi_\ell}(\sigma)$ and $\bar{\zeta_\ell}(\sigma) = s\, \bar{\psi_\ell}(\sigma)$, Eq.~\eqref{eq:WaveEq} is reexpressed an eigenvalue problem ${\rm L} {\rm u} = s {\rm u}$ for the operator \cite{Jaramillo:2020tuu}
\beq
\label{eq:QNM_EV}
{\rm L} = \left(
\begin{array}{cc}
0 & 1 \\
w^{-1}\boldsymbol {L_1} & w^{-1}\boldsymbol{ L_2}
\end{array}
\right),
\quad
{\rm u} = \left(
\begin{array}{c}
\bar\psi \\
\bar\zeta
\end{array}
\right) 
\eeq 
with the operators
\beq
\label{eq:L1}
\boldsymbol {L_1} &=& \dfrac{d}{d\sigma} \left( p(\sigma) \dfrac{d}{d\sigma}\right) - \bar V_\ell(\sigma), \\
\label{eq:L2}
\boldsymbol {L_2} &=& 2\gamma(\sigma) \dfrac{d}{d\sigma} + \gamma'(\sigma),
\eeq
and the rescaled potential $\bar V_{\ell}$ given by
\beq
\label{eq:hyp_pot}
\bar V_{\ell}(\sigma) = \dfrac{\lambda^2}{p(\sigma)} V_{\ell}(r).
\eeq

\section{Spacetime Limits}\label{sec:limits}
\begin{figure*}[ht!]
    \includegraphics{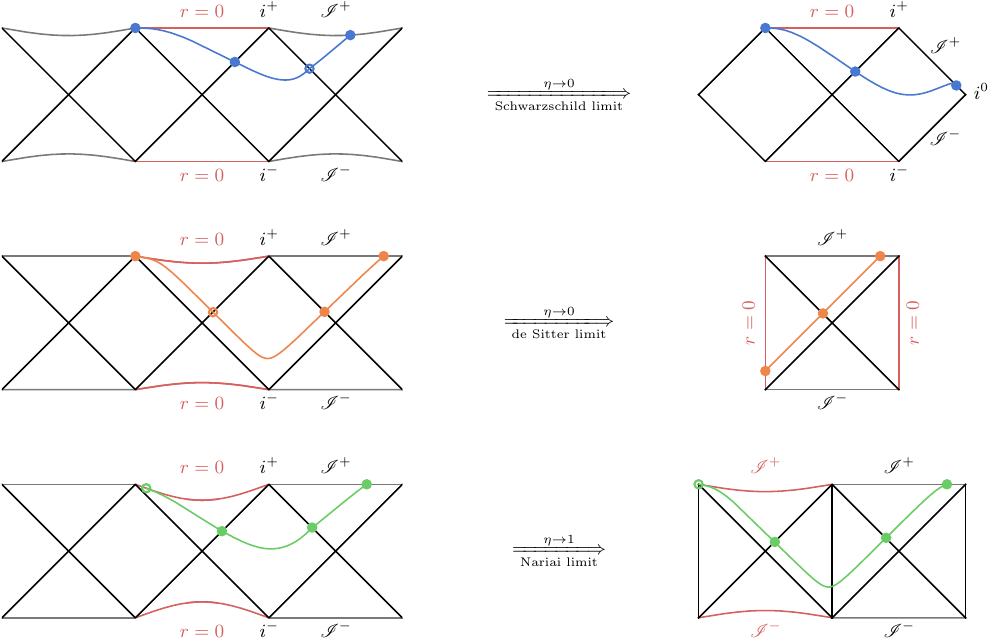}
    \caption{Penrose diagrams for the Schwarzschild--de Sitter spacetime and their respective limits achieved within the hyperboloidal framework as a natural implementation of Geroch’s geometric limiting procedure~\cite{Geroch:1969ca}. Solid points represent surfaces with a fixed coordinate value, whereas surfaces with an empty dot move freely in the grid. In all cases, $\scri^+$ is fixed at a coordinate location $\sigma =0$. Top: the Schwarzschild scenario. The black hole horizon fixes the spacetime length scale $\lambda = r_H$. The event horizon ${\cal H}^+$ and the singularity are fixed, but the cosmological horizon is free. The corresponding limit $\eta \rightarrow 0$ is the Schwarzschild spacetime, where the cosmological horizon degenerates into  $\scri^+$. Middle: the de Sitter scenario. The cosmological horizon fixes the spacetime length scale $\lambda = r_\Lambda$. The cosmological horizon ${\cal C}^+$ and the singularity are fixed, but the event horizon is free. The corresponding limit $\eta \rightarrow 0$ is the de Sitter spacetime, where the event horizon tends to the surface $r=0$. Bottom: the Nariai scenario. The cosmological and event horizons ${\cal H}^+$ are fixed, but the singularity moves freely. The spacetime length scale incorporates a singular behavior $\lambda = r_H/(1-\eta)$, but the corresponding limit $\eta \rightarrow 1$ is finite into the Nariai spacetime. }
    \label{fig:penrose_diagrams}
\end{figure*}

We now demonstrate how the hyperboloidal framework provides a clear geometrical strategy to study the limits of the SdS spacetime. In particular, we explore the freedom in Eq.~\eqref{eq:def_rho} given by the parameters $\rho_0$ and $\rho_1$ to fix relevant spacetime surfaces at constant coordinate values. As discussed in Sec.~\ref{sec:SdS}, the limit $\eta \to 0$ can either yield the Schwarzschild or de Sitter spacetimes, whereas $\eta \to 1$ leads to the extremal configurations. For $\eta \to 1$ we only consider the limit into the Nariai spacetime, as this scenario is the one leading to a well-defined QNMs eigenvalue problem in the frequency domain. 

Figure~\ref{fig:penrose_diagrams} summarizes and illustrates limiting procedures within the hyperboloidal framework as detailed in the following sections.

\subsection{The Schwarzschild scenario}\label{sec:Schwarzschild_limit}

In the Schwarzschild scenario, the parameter $\eta$ is understood as a small deviation from the Schwarzschild geometry. Therefore, the characteristic length scale of the spacetime is given by $\lambda =r_H$. Moreover, to recover the Schwarzschild geometry as $\eta\rightarrow 0$, one must ensure that the black hole horizon is at a fixed surface $\sigma_H$, independent of $\eta$. 

The most simple choice is to fix the horizon surface at $\sigma_H=1$, achieved trivially by a the radial transformation~\eqref{eq:def_rho} with parameters
\beq
\label{eq:radial_par_SchwarzLimit}
(\rho_0, \rho_1) = (1,0) \Rightarrow r = \dfrac{r_H}{\sigma}.
\eeq
Equation~\eqref{eq:radial_par_SchwarzLimit} maps the singularity $r=0$ to $\sigma_{\rm sing} \rightarrow \infty$, whereas the cosmological horizon depends on the spacetime parameter $\eta$ directly via $\sigma_\Lambda = \eta$. Therefore, the limit $\eta \rightarrow 0$ corresponds to having the cosmological horizon degenerating into future null infinity as $ \sigma_\Lambda \rightarrow 0 $.

This limit corresponds to a discontinuous change in the topology of future null infinity. While the surface $\sigma=0$ is spacelike for $\eta\neq 0$, it becomes null when $\eta=0$.

Despite the topology change, the dimensionless tortoise function has a well-defined limit as $\eta \rightarrow 0$. With the radial compactification~\eqref{eq:radial_par_SchwarzLimit}, the terms in Eq.~\eqref{eq:tort_x} read
\beq
\label{eq:x_H_Schwarzschild}
    x_H(\sigma) &=&
     \frac{ \left( 1 + \eta + \eta^2 \right) }
    { \left( 1- \eta  \right) \left( 2 \eta + 1 \right) }
    \ln \left| 1 - \sigma \right|, \\
    \label{eq:x_Lambda_Schwarzschild}
    x_\Lambda(\sigma) &=&
    - \frac{ \left( 1 + \eta + \eta^2 \right) }
    { \eta \left( 1 - \eta \right) \left( 2 + \eta \right) }
    \ln \left| \dfrac{\sigma}{\eta} - 1 \right|, \\
   \label{eq:x_Neg_Schwarzschild}
    x_o(\sigma) &=&
    \frac{ \left( 1 + \eta + \eta^2 \right) ( 1 + \eta ) }
    { \eta ( 2 \eta + 1 ) ( \eta + 2 ) }
    \ln \left| 1 + \sigma \dfrac{1+\eta}{\eta} \right|.
\eeq
The function $x_H(\sigma)$ is well behaved at $\eta=0$. Even though the functions $x_\Lambda(\sigma)$ and $x_o(\sigma)$ individually diverge as $\eta\rightarrow 0$, these divergences have exactly opposite signs, i.e.
\beq
    x_\Lambda(\sigma) &=& - \dfrac{1}{2\eta} \ln \left( \dfrac{\sigma}{\eta} \right) + {\cal O}(\eta^0), \\ 
    x_o(\sigma) &=& \dfrac{1}{2\eta} \left( \dfrac{\sigma}{\eta} \right) + {\cal O}(\eta^0).
\eeq 
Therefore the contribution from the sum $x_\Lambda(\sigma) + x_o(\sigma)$ is regular as $\eta \rightarrow 0$ and one recovers the Schwarzschild expression for the tortoise coordinate\cite{PanossoMacedo:2023qzp}
\beq
\lim_{\eta \rightarrow 0} x(\sigma) = \dfrac{1}{\sigma} - \ln(\sigma) + \ln(1-\sigma).
\eeq
The regularity of the combination $x_\Lambda(\sigma) + x_o(\sigma)$ plays a fundamental role when taking the limit $\eta\rightarrow 0$ within the hyperboloidal coordinate system. The height function defined in Eq.~\eqref{eq:H_def} via the out-in strategy has precisely a factor $-x_\Lambda(\sigma) - x_o(\sigma)$, which ensures a regular Schwarzschild limit into the expected minimal gauge expression in the Schwarzschild spacetime\cite{PanossoMacedo:2023qzp}
\beq
\lim_{\eta \rightarrow 0} H(\sigma) = -\dfrac{1}{\sigma} + \ln(\sigma) + \ln(1-\sigma).
\eeq
However, hyperboloidal foliations may arise from different choices of height functions $H(\sigma)$, some of which with aN ill-defined limit $\eta \rightarrow 0$. Indeed, as mentioned in Sec.~\ref{sec:hyp_coord}, the in-out strategy does provide a functioning hyperboloidal coordinate system when $\eta \neq 0$, but the combination $-x_\Lambda(\sigma) + x_o(\sigma) \sim \eta^{-1}$ in the height function has a singular limit $\eta \rightarrow 0$.

Equations~\eqref{eq:x_H_Schwarzschild}--\eqref{eq:x_Neg_Schwarzschild} provide all the necessary ingredients to formulate the QNM eigenvalue problem \eqref{eq:QNM_EV}. With $x_H(\sigma)$, $x_\Lambda(\sigma)$, and $x_o(\sigma)$, one constructs $x(\sigma)$ and $H(\sigma)$ in Eqs.~\eqref{eq:tort_x} and \eqref{eq:H_def}. From these quantities, the line elements \eqref{eq:hyp_line_func1} and \eqref{eq:hyp_line_func2}, as well as the rescaled potential \eqref{eq:hyp_pot} building up the operator in Eq.\eqref{eq:QNM_EV} follow directly. Having fixed the reference length scale to the horizon's size $\lambda = r_H$, the resulting QNM frequencies \eqref{eq:QNM_lambda} associated with the configuration appropriated to the Schwarzschild limit are expressed in terms of the dimensionless values
\beq
\label{eq:Sch_omega}
\omega^{\rm Sch} = r_H \omega.
\eeq

\subsection{The de Sitter scenario}\label{sec:de_Sitter_limit}

To study the de Sitter scenario, the cosmological horizon is the natural characteristic spacetime length scale $\lambda = r_\Lambda$. Since the parameter $\eta$ is understood as a small deviation from the de Sitter geometry, one must ensure that the cosmological horizon is fixed at surface $\sigma_\Lambda$, independent of $\kappa$. 

Similar to the previous section, a simple choice is to fix $\sigma_\Lambda = 1$, which could be easily achieved by a transformation $r = r_\Lambda /\sigma$. As before, this choice pushes the coordinate location of the surface $r=0$ into $\sigma_{\rm sing} \rightarrow \infty$. Apart from that, it also maps the black hole horizon into the surface $\sigma_H = \eta^{-1}$, which makes $ \sigma_H $ divergent in the limit $ \eta \rightarrow 0$. Such a divergence is consistent with our expectations. For $\eta\neq 0$, $r=0$ is a spacelike hypersurface corresponding to the BH singularity. As $\eta \rightarrow 0$, the black hole horizon degenerates into the singularity $r=0$ ($\sigma_{\rm sing} \rightarrow \infty$), and the surface $r=0$ changes topology, becoming a regular timelike hypersurface, representing the origin of the coordinate system.

However,  this radial compactification is not optimal for numerical studies, where the numerical domain is defined in the exterior BH region $\sigma \in [ \sigma_\Lambda, \sigma_H  ]$. As $\eta\rightarrow 0$, the domain stretches out with $\sigma_H$ assuming very high values.

To solve this issue, we use the freedom in Eq.~\eqref{eq:def_rho} to map the singularity $r=0$ into a fixed, but finite coordinate value $1<\sigma_{\rm sing}<\infty$. By imposing $r(\sigma_{\rm sing})=0$ and $r(1)=r_\Lambda$ into the radial Eqs.~\eqref{eq:hyp_coord} and \eqref{eq:def_rho}, one obtains
\beq
\label{eq:radial_trasfo_deSitter_Limit}
(\rho_0, \rho_1) &=&  \left( \dfrac{\sigma_{\rm sing}}{\sigma_{\rm sing}-1} , \dfrac{-1}{\sigma_{\rm sing}-1}\right) \nn \\
\Rightarrow r &=& \dfrac{r_\Lambda}{\sigma} \dfrac{\sigma_{\rm sing} - \sigma}{\sigma_{\rm sing} - 1}.
\eeq
In this way, the coordinate location of the event horizon becomes
\beq
\sigma_H = \dfrac{\sigma_{\rm sing}}{1 + \eta (\sigma_{\rm sing} - 1)},
\eeq
and as expected, $1< \sigma_H = \sigma_{\rm sing} < \infty$ when $\eta=0$. 

With Eq.~\eqref{eq:radial_trasfo_deSitter_Limit} substituted into \eqref{eq:rh_def} one reads the individual terms of the dimensionless tortoise coordinate as
\beq
    \label{eq:x_H_dS}
    x_H(\sigma) &=&
    \frac{ \eta \left( 1 + \eta + \eta^2 \right) }{ ( 1 + 2 \eta ) ( 1 - \eta ) }
    \ln \left| 1  -  \eta \sigma  \dfrac{\sigma_{\rm sing} - 1}{\sigma_{\rm sing} - \sigma} \right|, \\
    x_\Lambda(\sigma) &=&
    - \frac{ \left( 1 + \eta + \eta^2 \right) }{ ( \eta + 2 ) ( 1 - \eta ) }
    \ln \left| 1  -  \sigma  \dfrac{\sigma_{\rm sing} - 1}{\sigma_{\rm sing} - \sigma}  \right|, \\
        \label{eq:x_Neg_dS}
    x_o(\sigma) &=&
    \frac{(1+\eta)(1+\eta+\eta^2)}{ (2+\eta)(1+2\eta) } \nn \\
    && \times \ln \left| 1  +  (1+\eta) \sigma  \dfrac{\sigma_{\rm sing} - 1}{\sigma_{\rm sing} - \sigma} \right|.
\eeq
All these terns have a well-defined limit $\eta \rightarrow 0$, in particular, with $x_H(\sigma) \rightarrow 0$. Thus, the dimensionless tortoise coordinate and the height function in the de Sitter spacetime become
\beq
\label{eq:dsLimit_x_H}
\lim_{\eta \rightarrow 0} x(\sigma) =  - \lim_{\eta \rightarrow 0} H(\sigma) = \dfrac{1}{2}  \ln \left| \dfrac{\sigma_{\rm sing} + \sigma \left( \sigma_{\rm sing} -2\right) }{\sigma_{\rm sing} (\sigma -1)} \right|.
\eeq
Without loss of generality, we fix the parameter $\sigma_{\rm sing} = 2$, which simplifies the above results. 

Equation~\eqref{eq:dsLimit_x_H} shows that at $\eta =0$, the height function coincides with the tortoise function up to an overall minus sign. This result implies that the time hypersurfaces $\tau=$constant becomes an outgoing null coordinate as $\eta\rightarrow 0$, when the hyperboloidal coordinate is constructed within the out-in minimal gauge strategy according to Eq.~\eqref{eq:H_def}.

As in the previous section, Eqs.~\eqref{eq:x_H_dS}--\eqref{eq:x_Neg_dS} provide all the elements to formulate the QNM eigenvalue problem \eqref{eq:QNM_EV}. However, as a consequence of the null slices in the exact de Sitter limit $\eta =0$, the function $w(\sigma)$ vanishes altogether, whereas the conformal potential diverges as $(\sigma-\sigma_{\rm sing})^{-2}$ at the symmetry axis, cf.~definitions \eqref{eq:hyp_line_func2} and \eqref{eq:hyp_pot}. Thus, the QNM operator as in Eq.~\eqref{eq:QNM_EV} becomes ill defined, and the eigenvalue problem for the de Sitter spacetime must be reformulated into the generalized eigenvalue problem
\beq
\label{eq:EV_dS_Exact}
 \boldsymbol {\tilde L_1} \bar\psi = - s \boldsymbol {\tilde L_2} \bar\psi,
\eeq
with $ \boldsymbol {\tilde L_i} = (\sigma-\sigma_{\rm sing})^{2} \boldsymbol {L_i}$, for  $i=1,2$. We emphasize that the previous transformation and the generalized eigenvalue problem \eqref{eq:EV_dS_Exact} only applies in the exact de Sitter (dS) limit $\eta = 0$.

Furthermore, the setup appropriated to the dS limit has $\lambda = r_\Lambda$ as the reference length scale, so the resulting dimensionless QNM frequencies \eqref{eq:QNM_lambda} are
\beq
\label{eq:dS_omega}
\omega^{\rm dS} &=& r_\Lambda \omega \nn  \\
&=& \eta^{-1} \omega^{\rm Sch}.
\eeq

\subsection{The Nariai scenario}\label{sec:Nariai_limit}

Neither of the previous configurations are appropriate to study the extremal limit $\eta\rightarrow 1$. In both cases, Eqs.~\eqref{eq:x_H_Schwarzschild}--\eqref{eq:x_Neg_Schwarzschild} or Eqs.~\eqref{eq:x_H_dS}--\eqref{eq:x_Neg_dS} yield a line element \eqref{eq:conf_line_element} behaving as
\beq
\bar g_{\tau \sigma} \sim \left( 1- \eta \right)^{-1}, \quad  
\bar g_{\sigma \sigma} \sim \left( 1- \eta \right)^{-2}. 
\eeq
This result is not surprising, as it reflects the degeneracy of the horizon coordinate values $r_H = r_\Lambda = r_{\rm ext}$ with
\beq 
\label{eq:horz_ext}
r_{\rm ext} = 3M = \sqrt{\Lambda^{-1}}.
\eeq 
In the Schwarzschild scenario, the coordinate value for the cosmological horizon $\sigma_\Lambda(\eta)$ depends explicitly on the parameter $\eta$, with the two surfaces degenerating in the limit $\eta\rightarrow 1$, i.e., $\sigma_\Lambda(1) = \sigma_H = 1$. The same occurs in the de Sitter scenario, now with the black hole coordinate value $\sigma_H(\eta)$ having the explicit $\eta$ dependence. The limiting process shows the same degeneracy $\sigma_H(1) = \sigma_\Lambda = 1$. 

To properly obtain the spacetime in the extremal limit $\eta\rightarrow 1$, one must map the two horizons $ r_H $ and $ r_\Lambda $ into two distinct  hypersurfaces $\sigma_H\neq \sigma_\Lambda$, fixed at coordinate values independent of the parameter $\eta$. Without loss of generality, we can keep the event horizon at $\sigma_H = 1$ and fix the cosmological horizon at the value $\sigma_\Lambda = 1/2$. By imposing $r(1)=r_H$ and $r(1/2)=r_\Lambda$ for $\eta \neq 1$ into the radial equations~\eqref{eq:hyp_coord} and \eqref{eq:def_rho}, we obtain
\beq
\label{eq:rhol_trasfo_Ext_Limit}
&(\rho_0, \rho_1) =  \left( \dfrac{r_H(1-\eta)}{\lambda \eta} , -\dfrac{r_H(1-2\eta)}{\lambda \eta}\right), \\
\label{eq:radial_trasfo_Ext_Limit}
&r = \dfrac{r_H \bigg((1-\sigma) - \eta (1-2\sigma) \bigg)}{\eta \sigma} {}.
\eeq
In the limit $\eta\rightarrow 1$, the above transformation is actually singular since Eq.~\eqref{eq:radial_trasfo_Ext_Limit} reduces to $r(\sigma) = r_H$. This behavior is well known, and typical for obtaining the near-horizon geometry of extremal black holes~\cite{Carroll:2009maa,Kunduri:2013gce}. A regular spacetime arises once one combines the singular radial transformation \eqref{eq:radial_trasfo_Ext_Limit} with a singular map in the time coordinate $t\rightarrow t/(1-\eta)$. As we will show, the characteristic length scale $\lambda$ will incorporate the troublesome factor $(1-\eta)$. Indeed, Eq.~\eqref{eq:radial_trasfo_Ext_Limit} determines the individual terms of the dimensionless tortoise coordinate
\beq
    \label{eq:x_H_Ext}
    x_H(\sigma) &=&
    \dfrac{ r_H \left( 1 + \eta + \eta^2 \right) }{ \lambda ( 1 + 2\eta ) ( 1 - \eta ) }
    \ln \left| 1  -  \sigma \right|, \\
     x_\Lambda(\sigma) &=& -
    \dfrac{ r_H \left( 1 + \eta + \eta^2 \right) }{ \lambda \eta ( 2 + \eta ) ( 1 - \eta ) }
    \ln \left| 2 \sigma - 1\right|, \\
        \label{eq:x_Neg_Ext}
    x_o(\sigma) &=&
    \dfrac{ r_H (1+\eta) \left( 1 + \eta + \eta^2 \right) }{ \lambda \eta ( 1 + 2\eta ) ( 2 + \eta ) }
    \ln \left| 1  +  \dfrac{3 \eta \sigma}{1-\eta} \right|.
\eeq
As a consequence, the dimensionless tortoise coordinate and height function behave as
\beq
x &=& \dfrac{r_H}{\lambda (1-\eta)} \bigg( \ln \left| \dfrac{2 \sigma - 1}{1-\sigma}\right|  + {\cal O}(1- \eta)  \bigg), \\
H &=& \dfrac{r_H}{\lambda (1-\eta)} \bigg( \ln \left| (2 \sigma - 1)(1-\sigma)\right|  + {\cal O}(1- \eta)  \bigg).
\eeq
To ensure a well-behaved limit $\eta\rightarrow 1$, one must set the characteristic length scale to
\beq
\label{eq:length_ext}
\lambda = \dfrac{r_H}{1- \eta}.
\eeq 
A similar argument follows from the line element \eqref{eq:conf_line_element}, as its components behave as
\beq
\label{eq:gtautau_ext}
\bar g_{\tau \tau} &=& - \lambda^2 (1-\eta)^2 \bigg(  (1-\sigma)(2\sigma -1) + {\cal O }(1-\eta) \bigg), \\
\label{eq:gtausig_ext}
\bar g_{\tau \sigma} &=& \lambda r_H (1-\eta) \bigg(  (3-4\sigma)  \bigg), \\
\label{eq:gsigsig_ext}
\bar g_{\sigma \sigma} &=& 8 r_H^2 + {\cal O }(1-\eta).
\eeq
With Eq.~\eqref{eq:length_ext}, the limit $\eta\rightarrow 1$ yields a regular physical metric for the extremal SdS spacetime in hyperboloidal coordinates, where we identify the black hole horizon $r_H$ as the extreme horizon $r_E$,
\beq
\label{eq:Nariai}
\dfrac{d s^2}{ r_E^2} &=& - \dfrac{(1-\sigma)(2\sigma-1)}{\sigma^2} d\tau^2  \\ 
&& + \dfrac{2 (3-4\sigma)}{\sigma^2} d\sigma d\tau  + \dfrac{8}{\sigma^2} d\sigma^2 + d\varpi^2. \nn
\eeq
To verify that Eq.~\eqref{eq:Nariai} corresponds to the Nariai spacetime, we consider the Nariai line element in its traditional form~\cite{1950SRToh..34..160N}
\beq
\label{eq:Nariai_traditional}
ds_N^2 = - \dfrac{r_E^2 - r_N^2}{r_E^2} dt_N^2 + \dfrac{r_E^2}{(r_E^2 - r_N^2)} dr_N^2+ r_E^2 d\varpi^2.
\eeq
Following the strategy introduced in Sec.~\ref{sec:minimal gauge}, a hyperboloidal transformation is found by mapping the horizons $r=\{ r_E, -r_E\}$ into $\sigma=\{1/2, 1\}$ and fixing the characteristic length scale to $\lambda = r_E/2$. Explicitly, the coordinate transformation reads
\beq
\label{eq:Ext_to_Nariai}
\dfrac{t_N}{r_E} &=& \dfrac{1}{2}\bigg( \tau - h(\sigma) \bigg), \quad \dfrac{r_N}{r_E} = \dfrac{2}{\sigma} - 3, \nn \\
\quad h(\sigma) &=& \ln(1-\sigma) + \ln\left(\sigma - \dfrac{1}{2}\right),
\eeq
which indeed maps the line element \eqref{eq:Nariai_traditional} into Eq.~\eqref{eq:Nariai}.

Note that the limiting strategy outlined to derive Eq.~\eqref{eq:Nariai} provides all the necessary tools to calculate the SdS QNMs as the eigenvalue problem \eqref{eq:QNM_EV} up to $\eta = 1$. However, the choice for a characteristic length scale as Eq.~\eqref{eq:length_ext} implies that the dimensionless QNM frequencies in the extremal limiting scenario scale as
\beq
\label{eq:QNM_ExtLimit}
\omega^{\rm Ext} &=& \dfrac{r_H \omega}{1- \eta} \nn \\
&=& \dfrac{\omega^{\rm Sch}}{1-\eta}
\eeq
if compared to the SdS frequencies $\omega^{\rm Sch}$ corresponding to the Schwarzschild scenario. 

On the other hand, assuming the existence of finite QNM frequencies $\omega_N$ associated with the original Nariai spacetime \eqref{eq:Nariai_traditional}, the direct hyperboloidal map \eqref{eq:Ext_to_Nariai} yields
\beq
\label{eq:QNM_freq_Ext_to_Nariai}
\omega^{\rm Ext} = \dfrac{r_E}{2} \omega_N.
\eeq
Thus, a direct comparison between Eqs.~\eqref{eq:QNM_ExtLimit} and \eqref{eq:QNM_freq_Ext_to_Nariai} indicates that the ratio $\omega^{\rm Sch}/(1-\eta)$ must be finite in the limit $\eta \to 1$ for some family of SdS QNM $\omega^{\rm Sch}$.

The reasoning above relies on assuming the existence of finite QNM frequencies $\omega_N$. The argument is valid, as it is known that the frequencies $\omega_N$ correspond to those of a P\"oschl-Teller potential, see, e.g.,~\cite{Casals:2009zh}. Independent of their connection with the Nariai spacetime, QNMs associated with the P\"oschl-Teller potential are typically employed as toy models in gravitational wave physics, with particular recent discussion in the context of the hyperboloidal framework~\cite{Jaramillo:2020tuu}.

To connect all these different strategies, we solve here the QNM problem of the aforementioned spacetimes analytically, where the relation between the different parametrization of the Nariai spacetime arises when treating wave equation for the massless scalar field $\Psi(x)$ in the time domain.

Explicitly, the massless scalar field's wave equation expressed in the coordinates yielding Eq.~\eqref{eq:Nariai_traditional} reads with a standard separation of variables in terms of the spherical harmonics\footnote{Note the absence of the term $\sim 1/r$ in this decomposition.}  $ \Psi(t_N, r_N, \theta, \varphi) = \Psi_\ell(t_N, r_N) Y_{\ell m}(\theta, \varphi)$,
\beq
\label{eq:wav_eq_Nariai_traditional}
\Bigg\{ \left(1-\dfrac{r_N^2}{r_E^2} \right) \dfrac{\partial}{\partial r_N} \left[ \left(1-\dfrac{r_N^2}{r_E^2}\right) \dfrac{\partial}{\partial r_N} \right] \nn \\
- \dfrac{\partial^2}{\partial t_N^2} -l(l+1)\left(1-\dfrac{r_N^2}{r_E^2}\right) \Bigg\} \Psi_\ell = 0.
\eeq
The P\"oschl-Teller form arises when one reexpresses Eq.~\eqref{eq:wav_eq_Nariai_traditional} in terms of the Nariai tortoise coordinate (e.g.~\cite{Casals:2009zh})
\beq
\label{eq:Nariai_to_PT_form}
r^*_{N} = r_E \tanh^{-1}\left(\dfrac{r_N}{r_E}\right).
\eeq
The wave equation will then read
\beq
\label{eq:pde_PT_form}
\left( - \dfrac{\partial^2}{t_N^2} + \dfrac{\partial^2}{\partial r^*_{N}{}^2} - V_{\rm PT}(r^*_{N})  \right) \Psi_\ell = 0
\eeq
where the P\"oschl-Teller potential
\beq
V_{\rm PT}(\tilde X) =  \dfrac{r_E^{-2}\,U_0}{\cosh^2 (r^*_{N}/r_E)}, \quad U_0 = \ell(\ell+1)
\eeq
becomes evident.

With the relation between the wave equation in the Nariai spacetime and the P\"oschl-Teller equation established via a direct coordinate transformation, we proceed to relate these wave equations with the hyperboloidal framework.  Reference~\cite{Jaramillo:2020tuu} introduced a hyperboloidal coordinate $(T,X)$ for the P\"oschl-Teller wave equation via
\beq
\label{eq:PT_form_to_PT_sol}
T &=& \dfrac{t_N}{r_E} -  \ln\left( \cosh\left(\dfrac{r^*_N}{r_E}\right)\right), \nn \\
X &=& \tanh\left(\dfrac{r^*_N}{r_E}\right) = \dfrac{r_N}{r_E},
\eeq
leading to the wave equation
\beq
\label{eq:pde_PT_sol}
\bigg( -\partial_T^2 & - & 2X\partial_T\partial_X - \partial_T - 2X\partial_X \nn \\
 &+& (1-X^2)\partial_X^2 - U_0 \bigg)\Psi_\ell = 0.
\eeq
With a Fourier transformation in $T$, Eq.~\eqref{eq:pde_PT_sol} becomes a second-order linear ordinary differential equation that can be solved analytically with the Gaussian hypergeometric function, e.g. Ref.~\cite{Jaramillo:2020tuu}. Enforcing the hypergeometric series to be truncated into a polynomial to meet the regularity conditions of the QNM solutions yields the P\"oschl-Teller frequencies
\beq
\label{eq:PT_freq}
\omega_{\rm PT} =  \pm \dfrac{\sqrt{4U_0-1}}{2} + i \left( n + \dfrac{1}{2} \right).
\eeq
Relating the dimensionless hyperboloidal coordinate $T$ for the  P\"oschl-Teller equation with the dimensionful Nariai time $t_E$ fixes the Nariai QNMs frequencies trivially to
\beq
\label{eq:Nariai_freq}
\omega_N &=& r_E^{-1} \omega_{\rm PT}.
\eeq
There remains to confirm the consistency between the Nariai QNMs $\omega_N$ with the values $\omega^{\rm Ext}$ in Eqs.~\eqref{eq:QNM_ExtLimit} and \eqref{eq:QNM_freq_Ext_to_Nariai}. The wave equation expressed in terms of the SdS extremal spacetime \eqref{eq:Nariai} reads
\beq
\label{eq:wav_eq_Nariai}
\Bigg\{- 8\partial_\tau^2 & + & ( 2(3-4\sigma)\partial_\sigma - 4 )\partial_\tau - (1- \sigma)(1- 2\sigma)\partial_\sigma^2 \nn \\
 &-& (4\sigma-3)\partial_\sigma - \dfrac{l(l+1)}{\sigma^2} \Bigg\} \Psi_\ell = 0.
\eeq
Equations~\eqref{eq:pde_PT_sol} and \eqref{eq:wav_eq_Nariai} have a common feature. They both have the horizons $\pm r_E$ as regular singular points, i.e. $X=\{-1,1\}$ in the former and $\sigma = \{1/2,1\}$ in the latter. Equation~\eqref{eq:wav_eq_Nariai}, however, has another irregular singular point explicitly at the coordinate location $\sigma = 0$. This point is mapped to $X\to \infty$ in \eqref{eq:pde_PT_sol} via the radial compactification
\beq
\label{eq:radial_PT_NariaiExtLimit}
X = \dfrac{2}{\sigma} - 3, \quad \sigma = \dfrac{2}{X+3}. 
\eeq 
Hence, systematically combining the coordinate transformations \eqref{eq:Ext_to_Nariai}, \eqref{eq:Nariai_to_PT_form}, and \eqref{eq:PT_form_to_PT_sol} yields the relation between the hyperboloidal P\"oschl-Teller coordinate $T$ from Ref.~\cite{Jaramillo:2020tuu} and the hyperboloidal Nariai-extremal SdS limit $\tau$ via
\beq
\label{eq:Ext_to_PT_sol}
T = \dfrac{1}{2} \left( \tau -2\ln\sigma + \ln8 \right).
\eeq
Combined, Eqs.~\eqref{eq:radial_PT_NariaiExtLimit} and \eqref{eq:Ext_to_PT_sol} consistently map the hyperboloidal Nariai wave equation \eqref{eq:wav_eq_Nariai} and the hyperboloidal P\"oschl-Teller \eqref{eq:pde_PT_sol} wave equation into each other. Additionally, it is clear in Eq.~\eqref{eq:Ext_to_PT_sol} that the factor of $1/2$ in the time coordinate transformation leads to the same overall factor we observe between the SdS extremal frequencies and P\"oschl-Teller frequencies as in Eqs.~\eqref{eq:QNM_freq_Ext_to_Nariai} and \eqref{eq:Nariai_freq}. 

These exact values also serve as a benchmark against the numerically obtained QNMs. The next section introduces the numerical methods employed in this work

\begin{figure*}[th!]
    \includegraphics{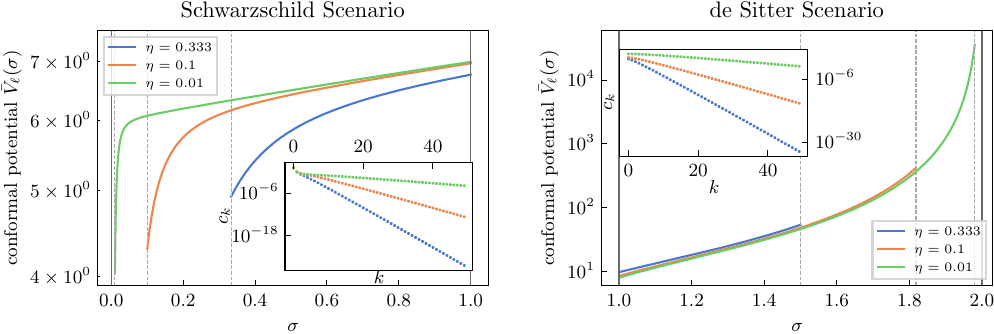}
    \caption{The conformal potential $\bar V_{\ell} ( \sigma )$ develops strong gradients around the horizons as $\eta \rightarrow 0$. In the Schwarzschild scenario (left), the strong gradient develops around the cosmological horizon as  $\sigma_\Lambda$ approaches future null infinity. In the de Sitter scenario (right), it develops around the event horizon as $\sigma_h$ approaches the singularity. Insets: the rate of convergence of the corresponding Chebyshev coefficients representing $\bar V_{\ell} ( \sigma )$ without AnMR. As $\eta \rightarrow 0$, the strong gradient near the horizons yields a loss of accuracy, which forces one to allocate much more computing power to get a proper accuracy for a QNM solver. }
    \label{fig:pot_limit}
\end{figure*}
\section{Numerical Methods}\label{sec:NumMeth}

To solve the QNM eigenvalue problem \eqref{eq:QNM_EV} we employ a collocation point spectral method having the Chebyshev polynomials as basis approximating the underlying functions. For that purpose, we fix a numerical resolution $N$ and introduce the Chebyshev-Lobbatto grid
\beq
\label{eq:grid}
\chi_i =\cos\left(\frac{\pi i}{N}\right) \ , \ i\in\{0,1,\ldots,N\} \ .
\eeq
parametrizing the domain $\chi \in [-1,1]$, where the Chebyshev polynomials of the first kind $T_k(\chi)$ are defined. By imposing that the approximated functions are exactly represented at the grid points \eqref{eq:grid}, one can approximate the derivative operator $\partial_\chi$ by the differentiation matrix
\beq
\label{e:CL_derivation_matrix}
\mathbb{D}^{ij}_{\chi} =
\left\{
\begin{array}{lcl}
  \displaystyle
  -\frac{2N^2+1}{6} & , & i=j=N , \\
  \displaystyle
  \frac{2N^2+1}{6} & , & i=j=0 , \\
  \displaystyle
  -\frac{\chi_j}{2(1-\chi_j)^2} & , & 0<i=j<N , \\
  \displaystyle
  \frac{\alpha_i}{\alpha_j} \frac{(-1)^{i-j}}{\chi_i-\chi_j} \ &,& i\neq j ,
\end{array}
\right.
\eeq
where
\beq
\alpha_i =
\left\{
\begin{array}{lcl}
  2 \ &,& \ i\in\{0,N\} , \\
  1 \ &,& \ i\in\{1,\ldots, N-1\} .
\end{array}
\right.
\eeq

The hyperboloidal radial coordinates, however, are defined in $\sigma \in[\sigma_\Lambda, \sigma_H]$. Typically, a linear map
\beq
\label{eq:sigma_chi}
\sigma(\chi) = \sigma_H \dfrac{1+\chi}{2} + \sigma_\Lambda \dfrac{1-\chi}{2}
\eeq
 $\sigma(\chi)$ from the spectral coordinate $\chi$ into $\sigma$ is employed, but as we will discuss, this choice is not ideal to explore the configuration in the limit $\eta \rightarrow 0$.

Indeed, the functions in the wave equations, such as the conformal potential $\bar V_{\ell}$, develop strong gradient around the domain boundaries as $\eta \rightarrow 0$. The left panel of Fig.~\ref{fig:pot_limit} displays the conformal potential $\bar V_{\ell}$ with Eq.~\eqref{eq:hyp_pot} calculated within the Schwarzschild scenario. Since future null infinity $\sigma = 0$ and the cosmological $\sigma_\Lambda(\eta)$ are close to each other as $\eta \rightarrow 0$, $\bar V_{\ell}$ develops strong gradients around $\sigma = \sigma_\Lambda$. The plot brings examples for the cases $\eta = 1/3, 1/10, \text{ and } 1/100 $ where the effect becomes visible. 

The inset shows the corresponding Chebyshev coefficients obtained when the linear map \eqref{eq:sigma_chi} is employed. One observes a significant loss of accuracy as $\eta\rightarrow 0$. An accurate numerical result for the QNMs would then require increasing the numerical truncation parameter $N$ to prohibitive high values. A similar effect happens also in the de Sitter scenario, as shown in the right panel of Fig.~\ref{fig:pot_limit}. In this case, however, strong gradients develop around the value $\sigma = \sigma_{\rm sing}$ (here $\sigma_{\rm sing}=2$) because the black hole horizon $\sigma_H(\eta)$ approaches $\sigma_{\rm sing}$ as $\eta \rightarrow 0$.

Thus, to enhance the numerical solver, we introduce the so-called AnMR, which we discuss in the next section.

\subsection{Analytical mesh refinement}
\label{sec:AnMR}

\begin{figure*}[t]
    \includegraphics{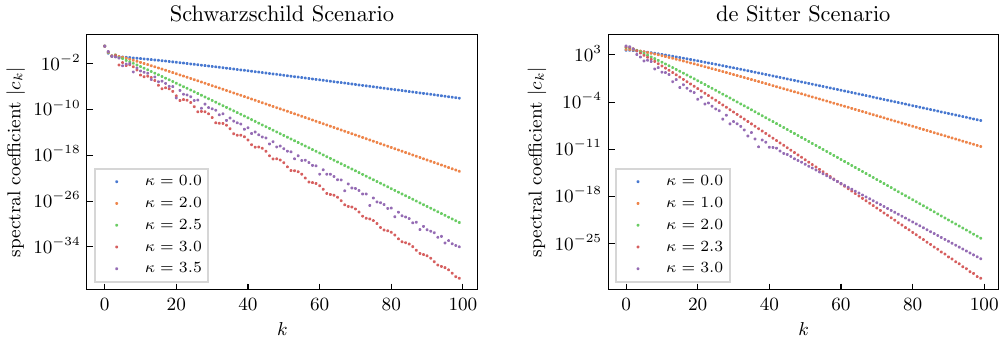}
    \caption{The optimization of the AnMR parameter $ \kappa $ based on the convergence rate of Chebyshev coefficients representing the conformal potential $\bar V_{\ell} ( \sigma )$. Both panels illustrate the case for $\eta = 1/100$. Different values of the AnMR parameter $ \kappa $ are tested numerically for the best accuracy in each case. The numerical test shows the optimized values roughly as $ \kappa \approx 3.0 $ for the Schwarzschild scenario (left). For the de Sitter scenario (right), the best value $ \kappa \approx 2.3 $ is qualitatively close $ \kappa =3 $, which allows one to assume the generic scaling \eqref{eq:bestAnMR}. }
    \label{fig:compare_kappa}
\end{figure*}

Instead of using the linear map \eqref{eq:sigma_chi}, we now consider the relation
\beq
\label{eq:chi_sigma_AnMR}
\sigma(\chi) = \sigma_H \dfrac{1+x(\chi)}{2} + \sigma_\Lambda \dfrac{1-x(\chi)}{2},
\eeq
with $x(\chi)$ the AnMR mapping the interval $[-1,1]$ into itself via
\beq
\label{eq:AnMR}
x = x_{\rm B}\Bigg( 1 - \dfrac{2 \sinh\left[ \kappa (1-x_{\rm B} \chi) \right] }{\sinh(2\kappa)} \Bigg),
\eeq
When $\kappa \rightarrow 0$, one recovers the identity $x(\chi) = \chi$. For $\kappa > 0$,  $x(\chi)$ accumulates the grid points toward the left boundary for $x_{\rm B} = -1$, or the right boundary for $x_{\rm B} = 1$. 

Figure~\ref{fig:compare_kappa} brings the Chebyshev coefficients associated with the conformal potential $\bar V_{\ell} ( \sigma )$ for the Schwarzschild scenario (left panel) and de Sitter scenario (right panel) for $\eta = 1/100$. Without the AnMR ($\kappa=0$), these coefficients are of order $c_k \sim 10^{-6}$ for a rather high numerical resolution $N=100$. As $\kappa$ increases and grid points accumulate around the region with steep gradients, we obtain an enhanced convergence rate up to an optimal value $\kappa_*$. For values $\kappa > \kappa_*$, the convergence rate gets worse. A systematic study of the parameter $\kappa$ for the Schwarzschild scenario yields the relation for the optimal value as
\beq
\label{eq:bestAnMR}
\kappa_* (\eta) = 1 - \ln \eta.
\eeq
In the de Sitter scenario, the optimal parameter $\kappa_*$ assumes values slightly bellow the relation in Eq.~\eqref{eq:bestAnMR}. However, as displayed in the right panel of Fig.~\ref{fig:compare_kappa} for $\eta = 1/100$, the results arising from \eqref{eq:bestAnMR} $\kappa = 3$ are qualitatively similar to the optimal value  $\kappa_* = 2.3$. Hence, one can also use the model \eqref{eq:bestAnMR} in the de Sitter scenario, and avoid a tedious optimization procedure for each individual $\eta$.

With the AnMR mappings \eqref{eq:chi_sigma_AnMR} and \eqref{eq:AnMR}, one then obtains a discrete representation for the derivative operator $\partial_\sigma$ via
    \beq
    & \mathbb{D}_{\sigma} = \overrightarrow{J^{-1}} \circ \mathbb{D}_{\chi}, \quad   \left(\overrightarrow{J^{-1}}\right)_i = \dfrac{1}{d \sigma(\chi_i)/d\chi}.
    \eeq
with the circle $\circ$ denoting the Hadamard (elementwise) product. Second-order derivatives follow directly from the product between two matrices $\mathbb{D}_{\sigma}$. With the discrete representation of the derivative operators, one can approximate the operators \eqref{eq:QNM_EV}--\eqref{eq:L2} via matrices. The QNM then correspond directly to the eigenvalues of the resulting matrix.

The ultimate test about the use of the AnMR for an efficient calculation of the QNMs is given by studying their numerical convergence.

\section{Quasinormal modes}\label{sec:Results}
In this section, we present our results for the QNMs of the SdS spacetime. Our numerical solver implementation that produces all the data in this section is made publicly available at \cite{zhou_2025_17218498}. We begin by examining the role of AnMR as an effective numerical technique for computing QNMs in extreme limiting regimes. To assess its performance, we perform a systematic convergence analysis. We then investigate the limiting behavior of the QNMs, separating the discussion into the two main families: the light ring (LR) modes and the dS modes.

Table~\ref{table:QNM_notation} summarizes the notation used to distinguish the behavior of these two families when analyzed in the geometrical scenarios most naturally adapted to the Schwarzschild, de Sitter, or extremal limits. As defined in Secs.~\ref{sec:Schwarzschild_limit} and \ref{sec:de_Sitter_limit}, superscripts such as in $\omega^{\rm Sch}$, $\omega^{\rm dS}$, and $\omega^{\rm Ext}$ denote the limiting geometrical configuration being considered, cf. Eqs.\eqref{eq:Sch_omega} and\eqref{eq:dS_omega}. Subscripts are used to label the particular family of QNMs---either the light ring modes $\omega_{\rm LR}$ or the de Sitter modes $\omega_{\rm dS}$.

All illustrative results are for the angular mode $\ell=2$, but the arguments and conclusions hold, in general.

\begin{table}[h]
    \label{table:QNM_notation}
    \caption{QNM families in different limiting scenarios}
    \begin{tabular}{| c || c | c |}
        \hline
          & Light ring       &  de Sitter        \\
          &  modes           &  modes         \\
        \hline
        Schwarzschild scenario   & $\omega^{\rm Sch}_{\rm LR}$        & $\omega^{\rm Sch}_{\rm dS}$  \\
        \hline
        de Sitter scenario    &    $\omega^{\rm dS}_{\rm LR}$      & $\omega^{\rm dS}_{\rm dS}$   \\
        \hline
        Nariai scenario    &    $\omega^{\rm Ext}_{\rm LR}$      & $\omega^{\rm Ext}_{\rm dS}$   \\
        \hline
    \end{tabular}    
\end{table}

\subsection{Convergence tests}
\begin{figure}[t!]
    \includegraphics{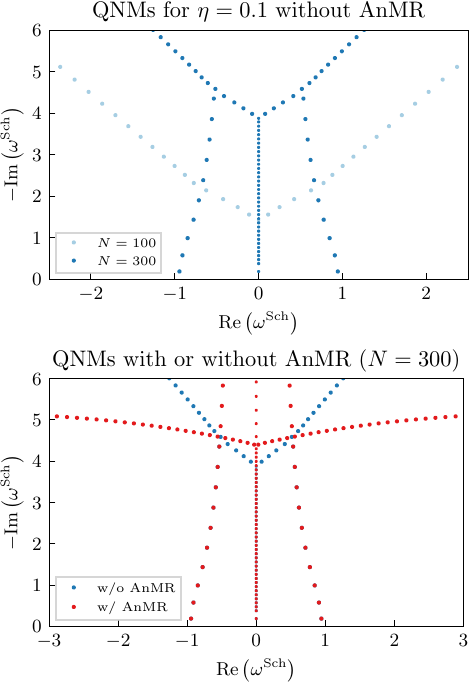}
    \caption{Top: light ring and the de Sitter QNM modes resolved without AnMR, under numerical resolutions of $N=100$ (light blue) and $N=300$ (dark blue). A noise branch in a {\textsf{V}} shape emanates from given critical value at the imaginary axis. Reliable data only exist under the noise branch, which increases with numerical resolution. Bottom: comparison of QNM values with or without AnMR (red and blue, respectively), for numerical resolution $N=300$. The noise branch changes its shape but the offset remains at same order of magnitudes.
    The AnMR technique allows us to calculate the light ring modes more accurately and we also find further values of light ring modes beyond the noise branch. The de Sitter modes are still valid only below the noise branch, but their evaluation is more precise, see Fig.~\ref{fig:conv_test}. }
    \label{fig:compare_AnMR_qualitative}
\end{figure}

The calculated solutions of the eigenvalue problem~\eqref{eq:QNM_EV} are shown in Fig.~\ref{fig:compare_AnMR_qualitative} for a configuration with a moderate value of $\eta = 0.1$. The results without AnMR are presented in the top panel, using numerical resolutions of $N = 100$ (light blue) and $N = 300$ (dark blue). In the figure, we identify two families of physical QNMs: the LR modes and the dS modes. Their values agree with those reported in the literature, up to a region bounded by a set of spurious data points, which we refer to as the “noise branch.” This noise branch originates at a critical point along the imaginary axis and spreads across the complex plane in a {\textsf{V}}-shaped pattern. Reliable QNM data are located below this branch. 

We observe that the critical value marking the onset of the noise branch increases with numerical resolution, thereby granting access to a larger portion of the physically relevant QNM spectrum. Additional tests at fixed resolution $N=100$ for varying $\eta$ further show that the onset of the noise branch also shifts to higher values as $\eta$ increases, confirming the growing difficulty of computing the QNM spectra in the limit $\eta\to0$.

When AnMR is enabled (bottom panel of Fig.~\ref{fig:compare_AnMR_qualitative}), the noise branch changes slightly in shape but remains at the same order of magnitude. Crucially, the method now enables us to compute data points above the noise branch. In particular, we are able to obtain additional physically relevant values for the LR modes. However, the purely imaginary points appearing above the noise branch do not represent an actual increase in correctly computed de Sitter modes: they fail to match theoretical predictions and do not converge with increasing numerical resolution $N$.

We finally recall that the appearance of spurious modes in the eigenvalue spectra (such as the above-mentioned noise branch) is a well-known feature of this type of numerical infrastructure. For instance, Ref.~\cite{Jaramillo:2020tuu} identifies not only modes along the imaginary axis---interpreted as a discrete representation of the branch cut in Schwarzschild spacetimes---but also additional spurious modes at large $|{\rm Im}(r_h \omega)|$ in the complex plane. When rotation is included~\cite{dePaula:2025fqt,Assaad:2025nbv}, spurious modes tend to cluster near the imaginary axis, whereas in systems modeling exotic compact objects~\cite{Boyanov:2022ark} a characteristic {\textsf{V}}-shaped distribution appears. The standard remedy is to compare spectra at two resolutions and filter out modes whose relative error exceeds a chosen tolerance~\cite{dePaula:2025fqt,Assaad:2025nbv,Siqueira:2025lww}.

Following the qualitative analysis, we proceed with a more detailed convergence study to assess the internal consistency of our numerical methods. To quantify how the computed QNM frequencies vary with numerical resolution, we define the highest resolution used, $N_{\rm ref}$, as the reference. Then, the QNM frequencies $\omega_n(N;\eta)$, computed at different resolutions $N$, are compared against the corresponding reference values $\omega_n(N_{\rm ref};\eta)$. The numerical error is defined as

\begin{equation}
\label{eq:QNM_error}
\varepsilon_n \left( N ; \eta \right) = \left| 1 - \frac{\omega_n \left( N ; \eta \right)}{\omega_n \left( N_\text{ref} ; \eta \right)} \right| ,
\end{equation}
where the index $n$ denotes the $n$th overtone. This error measure allows us to monitor the convergence behavior as a function of resolution. For the demonstration presented below, we consider scalar field perturbations with fixed angular momentum number $l = 2$. The maximum numerical resolution is set to $N = 300$.

\begin{figure}[t!]
    \includegraphics{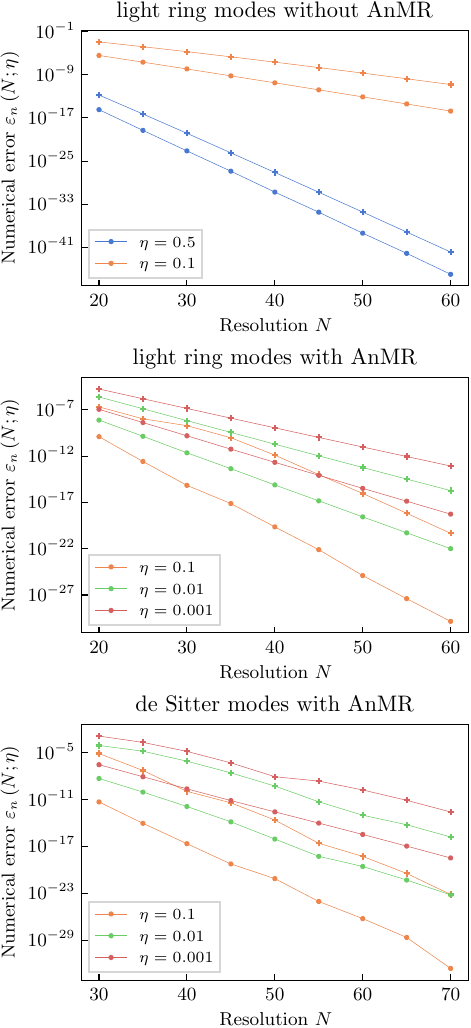}
    \caption{The convergence tests for QNMs, with dot markers representing the fundamental mode $n=0$, while plus markers denote the first overtone $n=1$. Color codes are kept for each value of $\eta$ across panels.
    Top: converge results for light ring modes without AnMR. Though exponential, the convergence rate reduces significantly as $\eta\rightarrow 0$.
    Middle: with AnMR enabled, the convergence rate increases and accurate results for smaller values $\eta$ are obtained with small numerical resources. 
    Bottom: AnMR also enhances the convergence rate for de Sitter modes in the limit $\eta \rightarrow 0$.
    }
    \label{fig:conv_test}
\end{figure}

We begin by analyzing the convergence behavior of the LR modes. The top panel of Fig.~\ref{fig:conv_test} shows the results obtained without AnMR. Here, dot markers correspond to the fundamental mode ($n = 0$), while plus markers represent the first overtone ($n = 1$). For $\eta = 0.5$ (blue), the convergence is very rapid: the numerical error drops by approximately 30 orders of magnitude as the resolution increases from $N = 20$ to $N = 60$. For $\eta = 0.1$ (orange), the convergence remains acceptable, with the error decreasing from $\sim 10^{-5}$ to $\sim 10^{-10}$. However, a clear degradation in the convergence rate is observed. As $\eta \rightarrow 0$, this trend continues: the convergence rate deteriorates further. Moreover, the noise branch begins to emerge near $\omega \sim 0$, contaminating the extraction of QNM overtones. As a result, achieving moderate accuracy for a meaningful number of QNMs becomes increasingly difficult and requires prohibitively high resolutions.

The middle panel of Fig.\ref{fig:conv_test} displays the convergence results with AnMR employed. In this case, the convergence rate for $\eta = 0.1$ (orange) significantly improves, reaching errors on the order of $\sim 10^{-30}$ near $N = 60$. This improvement also enables the computation of QNMs for smaller values of $\eta = 0.01$ (green) and $\eta = 0.001$ (red) with reliable convergence. Thus, AnMR makes it feasible to accurately study the regime $\eta \rightarrow 0$. A similar enhancement is observed for the de Sitter modes, whose results with AnMR are presented in the bottom panel of Fig.\ref{fig:conv_test}.

Even though these convergence tests were performed only within the setup adapted to the Schwarzschild limit, cf. Sec.~\ref{sec:Schwarzschild_limit}, the same conclusions are also valid in the case of the Sitter scenario. Indeed, for $\eta \neq 0$ the QNMs calculated in either setup are related to each other only by an overall rescaling factor as in Eq.~\eqref{eq:dS_omega}. Hence, the relative error \eqref{eq:QNM_error} remains unchanged.

With the geometry and numerics optimized to study the limiting cases of the SdS spacetime, we proceed to a comprehensive study of LR and dS modes in the limit $\eta\rightarrow 0$ and $\eta\rightarrow 1$ from the perspective of QNM spectra instability.

\subsection{Limits into Schwarzschild vs de Sitter spacetimes}
As discussed in Sec.~\ref{sec:limits}, the limiting spacetime obtained as $\eta \rightarrow 0$ depends on how the geometry is fixed via the hyperboloidal foliation. Depending on the chosen gauge, the limit may yield either the Schwarzschild or the de Sitter spacetime. The LR modes are characteristic QNMs of the former, while the dS modes are associated with the latter. In this subsection, we revisit the behavior of both LR and dS modes in the limit $\eta \rightarrow 0$, as we approach either the Schwarzschild spacetime or the de Sitter spacetime, interpreting the observed mode behavior within the perspective of QNM spectral instability.

\subsubsection{Light ring modes}

\begin{figure*}[ht!]
    \includegraphics{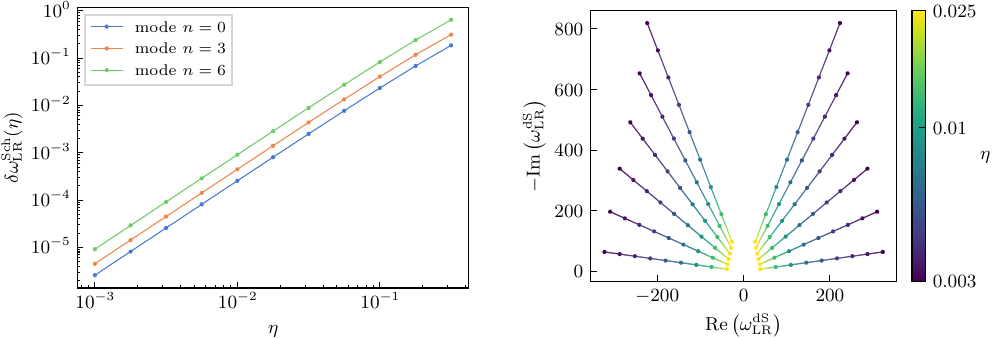}
    \caption{     
Left: relative deformation $\delta \omega^{\rm Schw}_{\rm LR}(\eta)$ of the LR QNMs in the Schwarzschild limiting scenario, computed at fixed numerical resolution $N = 300$. The deformation quantifies how the LR modes deviate from their Schwarzschild values as the cosmological constant is increased. The smooth, quadratic scaling $\delta \omega_n(\eta) \sim \eta^2$ confirms the stability of LR modes despite the change in the spacetime’s asymptotics structures.
Right: behavior of the LR QNMs in the extremal limit $\eta \to 0$ when the limiting procedure targets the de Sitter spacetime. When computed within the de Sitter scenario, the LR modes diverge as $\mathcal{O}(\eta^{-1})$, in agreement with the scaling predicted by Eq.~\eqref{eq:dS_omega}. This divergence illustrates a manifestation of QNM spectral instability: as a new small length scale $r_h \sim M$ is introduced, new QNM families emerge from infinity, reflecting changes in the causal structure near $r=0$. 
}
 \label{fig:deform_of_lr_modes}
\end{figure*}

In the Schwarzschild limiting scenario, LR modes converge to those of the pure Schwarzschild spacetime. This behavior reflects the expected stability of LR modes: small deviations from Schwarzschild geometry induced by the cosmological constant $\Lambda$ lead to correspondingly small deformations in the LR spectrum. 

In Fig.~\ref{fig:deform_of_lr_modes} (left panel), we display the smooth deformation of the LR modes as a function of the parameter $\eta$, which tracks the deviation from the Schwarzschild geometry within the hyperboloidal framework. The relative deformation is quantified by
\begin{equation}
\delta \omega^{\rm Schw}_{\rm LR}(\eta) = \frac{\left| \omega^{\rm Schw}_{\rm LR}{}(\eta) - \omega^{\rm Schw}_{\rm LR}{}(0) \right|}{\left| \omega^{\rm Schw}_{\rm LR}{}(0) \right|},
\end{equation}
where $\omega^{\rm Schw}_{\rm LR}{}(0)$ denotes the Schwarzschild modes.

Our numerical results reveal a clear scaling behavior: to leading order, the deformation of the LR modes follows a quadratic dependence on $\eta$, which translates into a linear dependence on the cosmological constant $\delta \omega_n(\eta) \sim \eta^2 \sim \Lambda$. This scaling confirms that the LR modes are robust under small cosmological deformations of the Schwarzschild background.

In contrast, the LR modes behave very differently when approaching the limit $\eta \to 0$ toward the de Sitter spacetime. According to the analytical estimate given in Eq.\eqref{eq:dS_omega}, these modes scale as $\mathcal{O}(\eta^{-1})$. This divergent behavior is illustrated in the right panel of Fig.~\ref{fig:deform_of_lr_modes}. 

While this scaling is well known in the literature, it also serves as an example of spectral instability: a new family of modes emerges from infinity as the introduction of a small length scale $r_H \sim M$ modifies the causal structure of the spacetime near the origin $r = 0$.

\subsubsection{de Sitter modes}

\begin{figure*}[htb]
    \includegraphics{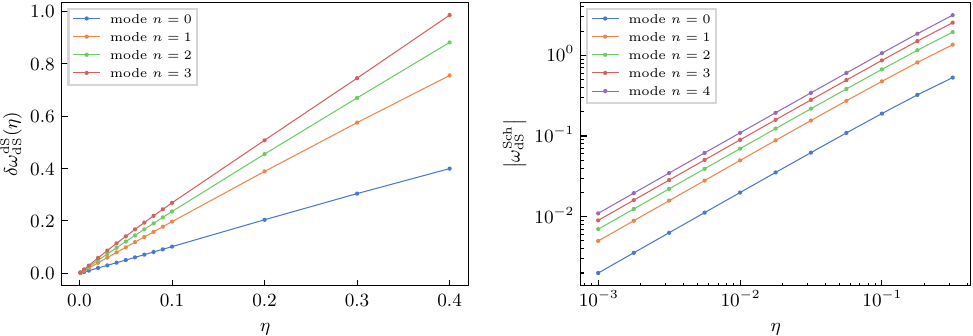}
    \caption{
        Left: the convergence of de Sitter modes in de Sitter limiting scenario as $ \eta \rightarrow 0 $. The y axis shows the deformation of de Sitter modes with $r_\Lambda$ as a typical length scale, which shows a smooth and linear relation $ \delta \omega^{\text{dS}}_{\text{dS}} \sim \eta $. This demonstrates the stability of dS modes in the de Sitter limiting scenario. Right: the equivalence of the left panel in Schwarzchild limiting scenario according to the relation in Eq.~\eqref{eq:dS_omega}. The de Sitter modes calculated in Schwarzschild limiting scenario $\omega^{\text{Sch}}_{\text{dS}}$ must shrink to 0 as $ \mathcal{O} ( \eta ) $ when $ \eta \rightarrow 0 $. Thus, the equivalent values calculated within the de Sitter scenario $\omega^{\text{dS}}_{\text{dS}}$ remain finite. }
    \label{fig:SdS_dS_modes}
\end{figure*}

The behavior of the dS modes follows a structure analogous to that of the LR modes, but with the notions of stability and instability reversed when comparing the Schwarzschild and de Sitter limiting scenarios. In the de Sitter scenario, the dS modes are stable as $\eta \sim 0$ corresponds to a small deviation from the pure de Sitter geometry.  In this regime, small values of $\eta$ induce only minor shifts in the dS QNM spectrum. 

This behavior is illustrated in the left panel of Fig.~\ref{fig:SdS_dS_modes}, where we display the relative difference
\beq
\delta \omega^{\rm dS}_{\rm dS}(\eta) = \dfrac{\omega^{\rm dS}_{\rm dS}(\eta) - \omega^{\rm dS}_{\rm dS}(0)}{\omega^{\rm dS}_{\rm dS}(0)},
\eeq
with $\omega^{\rm dS}_{\rm dS}(0)$ the exact QNMs values in the de Sitter spacetime.
As expected, we observe a quadratic scaling of the relative frequency shift, $\delta \omega^{\rm dS}_{\rm dS}(\eta) \sim \eta \sim M$, consistent with the stability of the dS spectrum under small perturbations near $r=0$ due to the emergence of a small black hole with mass $M \sim 0$.

To study the behavior of dS modes in the Schwarzschild limiting scenario, we invoke Eq.\eqref{eq:dS_omega} once more. In this case, taking the limit $\eta \to 0$ toward the Schwarzschild spacetime results in the dS mode frequencies scaling as $\omega \sim \mathcal{O}(\eta)$. That is, the dS modes collapse toward the origin of the complex frequency plane. This linear scaling is clearly visible in the right panel of Fig.\ref{fig:SdS_dS_modes}, where the dS modes approach $\omega = 0$ linearly as $\eta \to 0$.

This behavior also suggests a compelling interpretation within the framework of spectral instability. While the introduction of a small cosmological constant $\Lambda > 0$ has a negligible effect on the {\em discrete} spectrum of Schwarzschild spacetime (LR modes), it has a substantial impact on the {\em continuous} part of the spectrum (branch cut). Specifically, the small deviation $\eta \sim 0$ modifies the asymptotic structure of the spacetime, including a change in the topology of future null infinity $\scri^+$. As a consequence, the branch cut along the positive imaginary axis becomes fragmented. This branch cut is effectively replaced by the discrete set of de Sitter modes in the Schwarzschild--de Sitter spacetime. In this sense, the appearance of dS modes from the continuous Schwarzschild spectrum exemplifies a spectral instability triggered by the global geometric deformation of the spacetime.

An alternative perspective on the branch cut instability is to interpret it as the accumulation of discrete QNMs near $\omega = 0$ as $\eta \rightarrow 0$. We quantify this accumulation using a strategy inspired by the Weyl law for black holes proposed in Refs.~\cite{Jaramillo:2021tmt,Jaramillo:2022zvf,Besson:2024adi}, which consists of counting the number of modes within a specified region of the complex $\omega$ plane.

\begin{figure}[hb!]
    \includegraphics{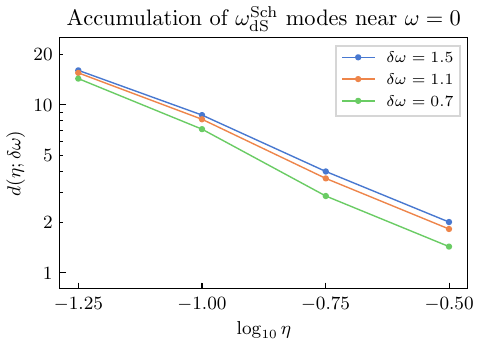}
    \caption{ Accumulation of de Sitter modes in the extremal limit $\eta \rightarrow 0$ when evaluated in the Schwarzschild limiting scenario. This accumulation is quantified by the QNM density $d(\eta; \delta \omega) \propto \eta^{-1}$, cf. Eq.\eqref{eq:density_of_qnms}. As $\eta \to 0$, the modes $\omega^{\rm Sch}_{\rm dS}$ accumulate infinitely at the origin. This divergence in QNM density supports the interpretation of the Schwarzschild branch cut as arising from an infinite accumulation of discrete de Sitter modes at $\omega = 0$. }
    \label{fig:accum_dSmodes}
\end{figure}

 \begin{figure*}[t!]
    \includegraphics{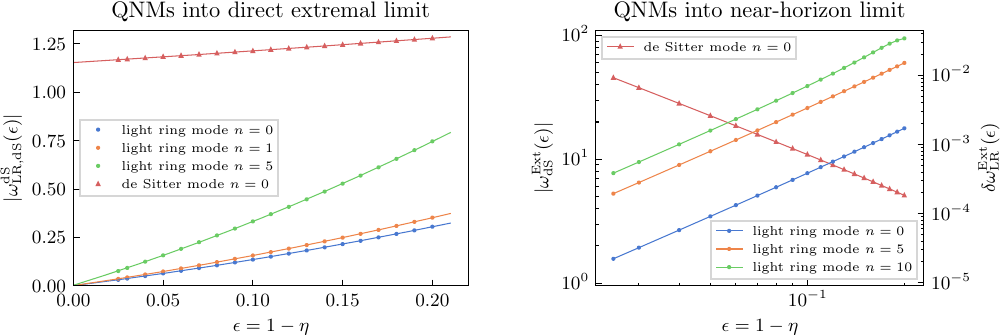}
    \caption{Extremality approach of $ | \omega^{\text{dS}} | \sim | \omega^{\text{Sch}} | $ as a function of $ \epsilon = 1 - \eta $. Left: the frequency-domain QNM problem is ill defined in the direct SdS extremal limit due to the absence of timelike Killing vector when $\epsilon=0$. For $\epsilon \gtrsim 0$, dS modes show a tendency of a nonvanishing constant, while LR modes approach $\omega = 0$. These tendencies are confirmed by a quadratic fit in the small parameter $\epsilon$. Right: the QNM problem is well defined for the near-horizon limit $\eta \to 1$ into the Nariai spacetime. The vanishing of the LR modes $\omega^{\text{dS}}_{\rm LR}$ in the Schwarzschild/de Sitter scenarios implies a finite limit $\omega^{\text{Ext}}_{\rm LR}$ into the exact P\"oschl-Teller values, whereas the dS modes $\omega^{\text{Ext}}_{\rm dS}$ disappear toward infinity, cf. \eqref{eq:QNM_ExtLimit}. }
    \label{fig:extlimit_from_dS}
\end{figure*}

To this end, we define a QNM density along a segment $\delta \omega = [a,b)$ of the imaginary axis as
\begin{equation}
d(\eta; \delta \omega) = \frac{\text{Number of QNMs in } \delta \omega}{\delta \omega}.
\label{eq:density_of_qnms}
\end{equation}
Since QNMs are discretely distributed, the density $d(\eta; \delta \omega)$ is only meaningful when the interval $\delta \omega$ is sufficiently large to include multiple modes.

Figure~\ref{fig:accum_dSmodes} shows the dependence of $d(\eta; \delta \omega)$ on $\eta$ for several values of $\delta \omega \gtrsim 0$, illustrating the trend toward an infinite accumulation of modes as $\eta \to 0$. 

To better understand the infinite accumulation of modes as $\delta \omega \to 0$, we consider a fixed number $M$ of modes located between two overtones, $\omega_n(\eta)$ and $\omega_{n+M}(\eta)$. By construction, the number of modes in this interval is $M$, and both end points scale as $\omega_n(\eta) \sim \mathcal{O}(\eta)$ and $\omega_{n+M}(\eta) \sim \mathcal{O}(\eta)$. Therefore, the length of the interval
$\delta\omega = \left| \omega_n(\eta) - \omega_{n+M}(\eta) \right|$
also scales as $\mathcal{O}(\eta)$, leading to a density that diverges as $d(\eta) \propto M/ \eta^{-1}$.

In the limit $\eta \rightarrow 0$, both end points of the segment collapse toward $\omega = 0$, implying an infinite accumulation of QNMs at the origin. This limiting behavior supports the interpretation of the Schwarzschild branch cut as emerging from a dense sequence of discrete de Sitter modes that accumulate at $\omega = 0$ as $\eta \to 0$.

\subsection{The extremal limit}

We conclude this section by analyzing the extremal limit of the SdS spacetime, corresponding to $\eta \to 1$. To track the approach to extremality, we introduce the small parameter
\begin{equation}
\epsilon = 1 - \eta,
\end{equation}
so that the extremal limit corresponds to $\epsilon \to 0$.

As discussed in Sec.\ref{sec:SdS}, the direct extremal SdS geometry lacks a global timelike Killing vector field. This absence precludes a well-defined QNM problem in the frequency domain and presents a fundamental obstruction to applying QNM eigenvalue solvers. In particular, the hyperboloidal formulation adapted to the Schwarzschild and de Sitter scenarios (Secs.\ref{sec:Schwarzschild_limit} and \ref{sec:de_Sitter_limit}) become degenerate in this limit, as the domain of dependence $\sigma \in [\sigma_\Lambda, \sigma_H]$ collapses to a point when $\sigma_\Lambda = \sigma_H$.

Despite this limitation, we can investigate the near-extremal regime with small but finite $\epsilon \gtrsim 0$. According to Eq.\eqref{eq:dS_omega}, both Schwarzschild and de Sitter scenarios yield equivalent QNM frequencies up to leading order: $\omega^{\rm dS} \approx \omega^{\rm Sch}$. The left panel of Fig.~\ref{fig:extlimit_from_dS} presents the behavior of LR and dS modes for $\epsilon \in [0.025, 0.2]$. The LR modes appear to shrink to zero, while the dS modes seem to converge to finite values.

To explore this trend more quantitatively, we perform a quadratic fit for each QNM family,
\begin{equation}
|\omega^{\rm dS}| = a_0 + a_1 \epsilon + a_2 \epsilon^2.
\end{equation}
The results confirm that $a_0 = 0$ for LR modes (within small fitting error fluctuation), implying that $\omega^{\rm Sch}_{\rm LR} = \mathcal{O}(\epsilon)$ as $\epsilon \to 0$, while $a_0 \neq 0$ for the dS modes, indicating a nonvanishing limit. This tendency provides a strong indication that the limit $\eta \to 1$ in Eq.~\eqref{eq:QNM_ExtLimit} is well defined for the LR family, which should correspond to the P\"oschl-Teller values as demonstrated in Sec.~\ref{sec:Nariai_limit}. 

Figure~\ref{fig:extlimit_from_dS} (right panel) illustrates this behavior explicitly, with QNMs computed directly within the Nariai scenario. In this setup, the hyperboloidal foliation spans the region between the horizons, fixed at $\sigma_\Lambda = 1/2$ and $\sigma_H=1$, and the time coordinate is properly rescaled to reach the exact Nariai spacetime as $\eta \to 1$. The plot shows the convergence of the LR modes $\omega^{\rm Ext}_{\rm LR}$ toward the analytic values, quantified on the right y axis by
\beq
\delta \omega^{\rm Ext}_{\rm LR}(\epsilon) = \dfrac{\omega^{\rm Ext}_{\rm LR}(\epsilon) - \omega^{\rm Ext}_{\rm LR}(0)}{\omega^{\rm Ext}_{\rm LR}(0)},
\eeq
with $\omega^{\rm Ext}_{\rm LR}(0)$ the Pöschl-Teller values, cf. Eqs.~\eqref{eq:PT_freq}, \eqref{eq:Nariai_freq}, and \eqref{eq:QNM_freq_Ext_to_Nariai}. In contrast, the dS modes diverge as $|\omega^{\rm Ext}_{\rm dS}| \propto \epsilon^{-1}$, a behavior read on the left y axis of Fig.~\ref{fig:extlimit_from_dS} (right panel).

At the exact extremal limit $\eta = 1$ ($\epsilon = 0$), a comparison with the Pöschl-Teller analytic solutions reveals remarkable numerical accuracy. For the fundamental LR mode, the relative error is below $10^{-155}$ when using high-precision arithmetic (with internal roundoff set to $10^{-160}$). Even the 50th overtone achieves a relative error as small as $10^{-28}$, demonstrating the robustness and precision of our numerical strategy in the extremal regime.

\section{Conclusion}\label{sec:Conclusion}
In this work, we revisited the QNM problem in the SdS spacetime using latest techniques in black hole perturbation theory, which allowed us to unify three fundamental aspects of the problem. From a geometrical perspective, the hyperboloidal framework provided a systematic way to implement limiting procedures for families of spacetimes, as originally formulated by Geroch~\cite{Geroch:1969ca}. From a numerical perspective, the QNM eigenvalue problem was solved using spectral methods based on the well-known Chebyshev polynomials, in contrast to previous approaches relying on the more intricate Bernstein polynomial basis. Finally, the results were interpreted through the lens of QNM spectral instability, offering new insights into how different families of QNMs behave under spacetime limits.
 
Specifically, Geroch showed that taking limits of a family of spacetimes is not uniquely defined~\cite{Geroch:1969ca}. For instance, when considering SdS as a one-parameter family characterized by $\eta = r_H/r_\Lambda$, the limit $\eta \to 0$ may yield either the Schwarzschild or de Sitter spacetime, depending on the identification of spacetime points. Similarly, the limit $\eta \to 1$ may lead to either the direct extremal SdS spacetime or its near-horizon geometry described by the Nariai solution. To define a meaningful limit, one must specify a prescription for identifying points across the family of spacetimes. This is precisely what the hyperboloidal framework enables: by fixing specific spacelike hypersurfaces to constant coordinate values, while maintaining a time foliation adapted to the asymptotic structure of the spacetime, the framework provides a natural and consistent procedure for implementing such limits in a manner best suited for the QNM problem. 

By controlling the geometrical limit, we observe that the (conformal) potential associated with the wave equation develops steep gradients in the parameter space near the limiting cases $\eta \to 0$. This feature, intrinsic to the physical properties of the problem, directly affects the performance of numerical schemes. In particular, spectral methods---whether based on Chebyshev or Bernstein polynomials---are sensitive to such gradients, leading to a deterioration in accuracy for small $\eta$  or requiring prohibitively high numerical resolution. To address this challenge, we demonstrate that the AnMR technique---successfully used in other settings~\cite{Meinel:2008kpy,Pynn:2016mtw,PanossoMacedo:2022fdi}---also proves highly effective in the context of QNM calculations. With AnMR, the eigenvalue solver is able to robustly recover in the limiting regimes both families of quasinormal modes in the SdS spacetime: the complex modes associated with the light ring (LR modes) and the purely imaginary modes characteristic of asymptotically de Sitter spacetimes (dS modes).

To explicitly implement Geroch’s strategy, the de Sitter spacetime is recovered in the limit $\eta \to 0$ by identifying the singularity at $r = 0$ and the cosmological horizon at $r_\Lambda$ with fixed values of the compactified radial coordinate, denoted $\sigma_{\rm sing}$ and $\sigma_\Lambda$, respectively. Meanwhile, the event horizon is allowed to vary with the parameter $\eta$. In this construction, the cosmological radius $r_\Lambda$ sets the fundamental length scale of the geometry, and as $\eta \to 0$, the coordinate location of the event horizon $\sigma_H$ merges with that of the singularity $\sigma_{\rm sing}$. Physically, this corresponds to a small black hole embedded in a nearly pure de Sitter background, where $\eta \ll 1$ modifies the causal structure around $r = 0$. In this regime, the de Sitter modes are found to be stable, smoothly deforming with $\eta$, and scaling as $\sim \eta$. In contrast, the LR modes can be interpreted as a manifestation of QNM spectral instability: the small parameter $\eta$ introduces a new physical length scale that triggers the appearance of a family of modes entering the complex frequency plane from infinity.
 
Similarly, the Schwarzschild spacetime is recovered in the limit $\eta \to 0$ by fixing the singularity at $r = 0$ and the event horizon at $r_H$ to constant values of the compactified radial coordinate, $\sigma_{\rm sing}$ and $\sigma_H$, respectively. In this scenario, it is the cosmological horizon that varies with $\eta$ and effectively recedes to future null infinity, $\scri^+$, as $\eta \to 0$. The event horizon radius $r_H$ thus sets the fundamental length scale of the geometry, and the small parameter $\eta \ll 1$ impacts the asymptotic structure of the spacetime---ultimately altering the nature of $\scri^+$ from a null into a spacelike surface.

It is important to note that not all hyperboloidal foliations yield a well-defined and smooth Schwarzschild limit. For instance, in the constructions proposed in Ref.~\cite{PanossoMacedo:2023qzp}, both the in-out and out-in strategies define minimal gauge hyperboloidal slicings for the SdS spacetime. However, only the out-in foliation admits a regular and controlled limit as $\eta \to 0$. 

In the Schwarzschild limiting scenario, the QNMs associated with the LR are stable, as they deform smoothly with respect to $\eta$, exhibiting a quadratic scaling $\sim \eta^2$. The behavior of the dS modes, however, reveals two distinct types of instability. Following the terminology of Ref.~\cite{Cheung:2021bol}, we first observe an overtaking instability, where the fundamental dS mode acquires a shorter decay timescale than the LR mode. More significantly, the Schwarzschild spacetime possesses a branch cut emerging from $\omega = 0$ and extending along the negative imaginary axis. As the asymptotic structure of the spacetime is altered by a small $\eta$, the new discrete family of purely imaginary QNMs (the dS modes) emerges from $\omega = 0$, effectively replacing---or ``destroying''---the original branch cut structure. 

An alternative interpretation of the branch cut is to view it as emerging from an infinite accumulation of discrete modes near $\omega = 0$ as $\eta \to 0$. Motivated by the Weyl law for black holes~\cite{Jaramillo:2021tmt,Jaramillo:2022zvf}, we introduced a heuristic measure of this effect by defining a QNM density along segments of the imaginary axis. This density captures the increasingly dense packing of purely imaginary modes as the Schwarzschild spacetime is approached in the limit $\eta \to 0$. However, a rigorous mathematical framework connecting the formation of a branch cut as a consequence of the accumulation of infinitely many discrete modes remains an open question.

Hence, this work opens a new avenue for exploring the notion of branch cut instability. As shown, in the Schwarzschild limiting scenario, a small $\eta \sim 0$ breaks the continuous branch cut into a discrete family of de Sitter modes. In other studies (e.g.,~\cite{Boyanov:2022ark}), {\em ad hoc} modifications to the wave equation potential have been observed to split the branch cut into {\textsf{V}}-shaped structures emerging from $\omega = 0$. Likewise, the inclusion of rotation in the spacetime~\cite{dePaula:2025fqt,Assaad:2025nbv} introduces features near the imaginary axis, often interpreted as numerical noise. While such features may indeed stem from the particular numerical strategy employed, these observations raise the intriguing possibility that branch cut instabilities---analogous to QNM instabilities---could carry genuine physical significance. This question is particularly relevant in the context of the late-time behavior of gravitational wave signals, where subtle spectral features might leave observable imprints.

The geometrical and numerical infrastructure developed here also proves robust in studying limiting scenarios approaching extremality. In the case of the SdS spacetime, the direct extremal limit lacks a timelike Killing vector, rendering a standard QNM analysis in the frequency domain ill defined. In practical terms, the QNM solver domains of dependence $\sigma \in [\sigma_\Lambda, \sigma_H]$ collapse into a single point $\sigma_H=\sigma_\Lambda$. However, a well-defined near-horizon description exists in the form of the Nariai spacetime. This limit can be naturally implemented within the hyperboloidal framework by fixing the event and cosmological horizons at {\em distinct} constant values $\sigma_H$ and $\sigma_\Lambda$, while rescaling the spacetime fundamental length scale with the diverging factor $\sim r_H / (1 - \eta)$. Within this construction, we systematically recover the correspondence between the QNMs in the Nariai geometry and those of the Pöschl-Teller potential. Moreover, we also provided the appropriate link with alternative hyperboloidal foliations of the Pöschl-Teller wave equation as discussed in~\cite{Jaramillo:2020tuu}.

Overall, the geometrical and numerical framework presented in this work lays a solid foundation for the analysis of spacetime limits in black hole perturbation theory. Its modular structure---based on hyperboloidal foliations, coordinate control of limiting procedures, and spectral methods enhanced by analytical mesh refinement---can be readily extended to more intricate spacetimes. In particular, configurations such as Reissner--Nordström--de Sitter, Kerr--de Sitter, or even Kerr--Newman--de Sitter spacetimes fall naturally within the scope of future investigations.

\acknowledgments

The Center of Gravity is a Center of Excellence funded by the Danish National Research Foundation under Grant No. 184.
R.P.M. acknowledges support by VILLUM Foundation (Grant No. VIL37766) and the DNRF Chair program (Grant No. DNRF162) by the Danish National Research Foundation.
This project also received financial support provided under the European Union’s H2020 ERC Advanced Grant “Black holes: Gravitational engines of discovery” Grant Agreement No. Gravitas--101052587, as well under the research and innovation program Marie Sklodowska-Curie Grant Agreements No. 101007855 and No. 101131233.
Views and opinions expressed are, however, those of the author only and do not necessarily reflect those of the European Union or the European Research Council. Neither the European Union nor the granting authority can be held responsible for them.

\bibliography{reference.bib}

\appendix

\end{document}